\documentclass[aip,
  rsi,amsmath,amssymb,
reprint,%
author-numerical,%
]{revtex4-1}

\usepackage{graphicx}
\usepackage{dcolumn}
\usepackage{bm}
\usepackage{comment}
\usepackage{titlesec}
\usepackage{lipsum}
\usepackage{textcomp}
\usepackage{gensymb}
\usepackage{color,colordvi}
\usepackage{ulem}
\usepackage{footmisc}
\begin{document}

\definecolor{purple}{rgb}{0.59,0.44,0.84}
\newcommand{\an}[1]{\textcolor[rgb]{0,0.5,0}{AN: \textbf{\textit{#1}}}}
\newcommand{\gn}[1]{{\textcolor{purple}{NG: #1}}}
\newcommand{\fc}[1]{{\textcolor{blue}{#1}}}
\newcommand{\ez}[1]{{\textcolor{magenta}{EZ: #1}}}


\title{Dynamical properties of different models of elastic polymer rings: confirming the link between deformation and fragility}


\author{Nicoletta Gnan}
\affiliation{CNR Institute for Complex Systems, Uos Sapienza, Piazzale Aldo Moro 2, 00185 Roma, Italy}
\affiliation{Department of Physics, Sapienza University of Rome, Piazzale Aldo Moro 2, 00185 Roma, Italy}
\email{nicoletta.gnan@cnr.it}
\author{Fabrizio Camerin}
\affiliation{CNR Institute for Complex Systems, Uos Sapienza, Piazzale Aldo Moro 2, 00185 Roma, Italy}
\affiliation{Department of Basic and Applied Sciences for Engineering, "Sapienza" University of Rome, via  Antonio Scarpa 14, 00161 Roma, Italy}

\author{Giovanni Del Monte}
\affiliation{Department of Physics, "Sapienza" University of Rome, Piazzale Aldo Moro 2, 00185 Roma, Italy}
\affiliation{CNR Institute for Complex Systems, Uos Sapienza, Piazzale Aldo Moro 2, 00185 Roma, Italy}
\affiliation{Center for Life NanoScience, Italian Institute of Technology, Viale Regina Elena 291, 00161 Roma, Italy}

\author{Andrea Ninarello}
\affiliation{CNR Institute for Complex Systems, Uos Sapienza, Piazzale Aldo Moro 2, 00185 Roma, Italy}
\affiliation{Department of Physics, Sapienza University of Rome, Piazzale Aldo Moro 2, 00185 Roma, Italy}

\author{Emanuela Zaccarelli}
\affiliation{CNR Institute for Complex Systems, Uos Sapienza, Piazzale Aldo Moro 2, 00185 Roma, Italy}
\affiliation{Department of Physics, Sapienza University of Rome, Piazzale Aldo Moro 2, 00185 Roma, Italy}

\date{\today}

\begin{abstract}
We report extensive numerical simulations of different models of 2D polymer rings with internal elasticity. We monitor the dynamical behavior of the rings as a function of the packing fraction, to address the effects of particle deformation on the collective response of the system. In particular, we compare three different models: (i) a recently investigated model [Gnan \& Zaccarelli, Nat. Phys. 15, 683 (2019)], where an inner hertzian field  providing the internal elasticity acts on the monomers of the ring, (ii) the same model where the effect of such a field on the center of mass is balanced by opposite forces and (iii) a semi-flexible model where an angular potential between adjacent monomers induces strong particle deformations. By analyzing the dynamics of the three models, we find that, in all cases, there exists a direct link between the system fragility and particle asphericity. Among the three, only the first model displays anomalous dynamics in the form of a super-diffusive behavior of the mean squared displacement and of a compressed exponential relaxation of the density auto-correlation function. We show that this is due to the combination of internal elasticity and the out-of-equilibrium force self-generated by each ring, both of which are necessary ingredients to induce such peculiar behavior often observed in experiments of colloidal gels. 
These findings reinforce the role of particle deformation, connected to internal elasticity, in driving the dynamical response of dense soft particles.
\end{abstract}

\maketitle

\section{Introduction}
Unlike hard spheres, several colloidal particles employed in soft matter physics, such as microgels, micelles, emulsions and star polymers, possess complex internal degrees of freedom, which provide them with an internal ``softness''~\cite{Doi}.
This property can be defined as the ratio between the single-particle elastic energy and the thermal energy~\cite{Vlassopoulos2012} and it manifests as the ability of soft particles to change both their volume and shape, by shrinking/swelling and deforming, thus affecting the mechanical and dynamical response as well as the phase behavior of the bulk suspensions they form. These effects become more and more important at high enough packing fractions where particles come into contact with each other.~\cite{Vlassopoulos2012,Vlassopoulos2014} 

Softness was also shown to have a prominent influence on effective interactions, both in microgels~\cite{Rossi2015,Bergman2018,Rovigatti2019,camerin2020microgels} and star polymers.~\cite{Jusufi1999,Jusufi2001}
From a dynamical perspective, the presence of an inverse correlation between softness and dynamical fragility in the supercooled regime
was put forward some years ago,~\cite{Mattsson2009, Schweizer2010} 
and it is still a very debated issue both in experiments~\cite{Nigro2017,Philippe2018} and in simulations~\cite{Sprakel2017,Higler2018}.
In addition, soft colloids can explore a variety of high density states, above the so-called jamming point~\cite{Zhang2009,Scheffold2014,Gury2019}, here loosely defined as the limit condition in which particles can fill the available space without deforming.
Under these conditions, the internal degrees of freedom of the particles give rise to effects like interpenetration~\cite{Mohanty2017}, shrinking and faceting~\cite{de2017deswelling,Conley2019,scheffold2020pathways}, which dominate the unusual dynamics and rheological behavior of these particles.~\cite{Vlassopoulos2012}

To model soft objects, one usually relies on the use of simple coarse-grained models; for instance in the case of microgels, a widely-employed model is the Hertz potential~\cite{Zhang2009,Berthier2010aa,Mohanty_2014,Bergman2018}, although it was recently shown that this can be considered to be valid only in the fluid regime~\cite{Rovigatti2019}.  Indeed, such simple models obviously neglect the polymeric nature of particles and some studies have started to consider modifications that allow to model some of the specific degrees of freedom of the particles~\cite{Claudio2009, GnanMacro2017, Moreno2018, rovigatti_review}. In particular, Urich and Denton~\cite{Urich_2016} have proposed to use a generalized hertzian model that accounts for an isotropic change of particle volume, thus being able to model shrinking effects. An extension of this model, incorporating dynamical behavior and particle polydispersity, has been recently investigated by Baul and Dzubiella~\cite{baul2020structure}.  A notable effort was also carried out by Higler and Sprakel, who investigated the dynamics of isotropically-deswelling particles, finding no dependence of the fragility on their internal softness~\cite{Higler2018}. In the need to go one step further and to explicitly introduce particle deformation, some of us recently proposed a new model of so-called Elastic Polymer Rings (EPR) in 2D~\cite{Gnan}, that incorporates both the polymeric and elastic features typical of soft spheres in 3D, but allows, at the same time, for a more efficient exploration, in terms of computing time, of the high density regime. This simple model was shown to capture shrinking, deformation and faceting at high densities. It was further characterized by the emergence of anomalous dynamics in the form of a compressed exponential behavior of the density auto-correlation function and of a super-diffusive behavior of the mean-squared displacement. In that case, one of the main findings was the identification of a clear link in simulations between particle deformation and fragility {  (which characterizes how fast the dynamics changes on increasing the packing fraction), with softer rings displaying a small fragility (strong systems) compared to stiffer rings (known as fragile systems)}, reinforcing the hypothesis that single-particle elasticity may dictate the collective behavior at large concentrations~\cite{Mattsson2009,Sprakel2017}.

In the original EPR model of Ref.~\cite{Gnan}, each ring was modeled as a bead-spring polymer connected in a circular shape and the internal elasticity was provided by an inner hertzian field.  The latter term represents, in a coarse-grained fashion, the elasticity of an underlying network in good solvent conditions, to mimic for example microgels, dendrimers or similar soft colloids. The hertzian field was chosen to act between each monomer of the ring and a reference point coinciding with the ring's center of mass, which amounts to consider the inner part of the ring as a coarse-grained object,  including network and solvent degrees of freedom. This inner field acts against the shrinking or the compression of the ring and tends to restore the maximally swollen condition. When the ring is symmetric (with at least 2-fold rotational symmetries), the inner field is strictly zero. However, as soon as the ring is slightly deformed, becoming asymmetric, for example in the presence of thermal fluctuations,  a non-zero net force, called hertzian force $\vec{F}_{H}$, acts on its center of mass. To restore a zero net force on the center of mass (force conservation), a contribution equal and opposite to $\vec{F}_{H}$ should be redistributed on all the monomers of the ring. This effectively leads to a reduction of the overall tendency of the system to deform at high densities.. In Ref.~\cite{Gnan}, such a restoring force was not employed, so that we can consider the original EPR model to be off-equilibrium and under the influence of the hertzian force that depends on the degrees of freedom of the deformed ring. Thus, $\vec{F}_{H}$ is self-generated in the system and increases with particle density.  For this reason, contrarily to what stated in Ref.~\cite{Gnan}, such EPR model cannot be considered as an `equilibrium' system due to this unbalanced force~\footnote{\label{note1} A correction statement is added to Ref.~\cite{Gnan} to notify the non-equilibrium aspects of the EPR model}.
To dissipate such a force, the use of a Langevin thermostat is thus necessary, because, on one hand, it mimics more realistically the microscopic dynamics of soft colloidal suspensions and, on the other hand, it ensures the dissipation of the center-of-mass force through the surrounding implicit solvent.  

In this work we  compare the EPR model with its  equilibrium version, here named equilibrium EPR (eq-EPR)  where, at each instant, the { center-of-mass} force acting on each ring is balanced by imposing an opposite force to all monomers constituting the ring. Under these conditions, the rings lose most of their ability to deform  at high densities and are subjected to a dramatic slowdown without displaying an anomalous behavior of the dynamics. Since in Ref.~\cite{Gnan}  a crucial effect of the deformation on the dynamics was demonstrated, we now take into account an additional model where particle deformation is enhanced in equilibrium. To this aim, we consider 2D semi-flexible polymer rings (SFPR), where the bead-spring model is complemented by an angular function which favors elongated arrangement of consecutive monomers, the latter playing the role of an effective elasticity. 
Similar models are commonly used in the literature to study ring polymers in 3D~\cite{Halverson_2011_I, Halverson_2011_II, Narros_2013, Michieletto_2016}, while our model is conceived to be a 2D schematic version  of complex, polymeric particles. We show that this system, although having only pairwise and three-body interactions, preserves much of the features of the original EPR model, including a reentrant dynamics and a strong tendency to deform. We investigate the dynamical behavior of the three models and, in particular, analyze the high density reentrant behavior that we observe. Similarly { to what was done} in Ref.~\cite{Gnan}, we build a modified Angell plot for the different models  and extract an effective fragility. We then find that a linear relation between fragility and particle asphericity holds for all investigated models, spanning a wide range of fragilities.

Surprisingly, we find that the SFPR model displays a much greater tendency to shape fluctuations with respect to the EPR system, thus continuously releasing the stress. This is, however, not sufficient to generate a  super-diffusive dynamics or a compressed relaxation, against the common assumption that stress release is responsible for such fast dynamics. It appears that the stress must be propagated, involving a collective response of the system that is found in the EPR model and not in the other models. To shed light on whether the out-of-equilibrium hertzian force is the only { cause} of the super-diffusive behavior, we additionally investigate the dynamics of a modified, non-equilibrium version of hertzian disks, where a force acting on the center-of-mass is added through the overlap between neighboring particles. Thus, this model can be considered conceptually analogous to the EPR model, but without the inclusion of particle elasticity, since disks lack internal degrees of freedom. We find that such out-of-equilibrium hertzian disks also do not display a super-diffusive behavior. This suggests that elasticity and off-equilibrium behavior are both necessary conditions to observe anomalous dynamics in soft systems.

 

The paper is organized as follows: in section \ref{sec:modandmeth} we define the different models and describe the numerical methods and the calculated observables. Next, in section \ref{sec:results} we investigate { the role played by} the hertzian force and how it influences the dynamical response of the system compared to the equilibrium models. We also report an analysis of particle deformation, focusing on the asphericity distributions and on the link with dynamical response, finding a direct relation between deformation and fragility, thus generalizing the results of Ref.~\cite{Gnan} for soft particles with internal elasticity. Finally, we discuss the role played by  the out-of-equilibrium hertzian force into the emergence of the super-diffusive motion and to this aim we compare  the results with the out-of-equilibrium hertzian disks showing that an interplay between elasticity and non-equilibrium is mandatory in {these} models for observing the super-diffusive dynamics.

\section{Models and Methods}\label{sec:modandmeth}
We investigate via extensive Langevin Dynamics simulations three different models of elastic polymer rings in 2D, that are based on the classical bead-spring model for polymers~\cite{Grest:1986aa}. Specifically, we 
consider $N=1000$ rings, each composed of $N_m=10$ monomers, interacting
among themselves via a Weeks-Chandler-Andersen
(WCA) repulsive term plus a Finitely-Extensible-Nonlinear-Elastic (FENE) potential acting only
among connected monomers, as
\begin{equation}
\label{eq:wca}
V_{\text{WCA}}(r)  =  
\begin{cases}
4\epsilon\left[\left(\frac{\sigma_m}{r}\right)^{12}-\left(\frac{\sigma_m}{r}\right)^6\right]+\epsilon & \quad \text{if} \quad r \le 2^{1/6}\sigma_m  \\
0 & \quad \text{if}  \quad  r > 2^{1/6}\sigma_m
\end{cases}
\end{equation}
and
\begin{equation}
V_{\text{FENE}}(r)  = 
-\epsilon k_F{l_0}^2\log\left[1-{\left(\frac{r}{l_0\sigma_m}\right)}^2\right], \quad r < l_0\sigma_m.
\label{eq:fene}
\end{equation}
Here $\epsilon$ sets the energy scale, $\sigma_m$ is the single monomer diameter, while $k_F=15,l_0=1.5$ are constants indicating the bond stiffness and the maximum extension of the polymer bond, respectively.

We employ a size polydispersity $\delta=12\%$ both for the monomer and for the ring size,
according to a log-normal distribution, to avoid crystallization at high packing fractions. The radius of each ring at infinite dilution, defined as the distance between each monomer and the center of mass in a circular shape, is $R_H=1.554\sigma_m$. 
We work at different nominal packing (area) fractions $\zeta=\frac{\pi}{4}\sum_{i=1}^N\sigma_{i,ring}^2/L^2$, where 
 $L$ is { size of} the box side and $\sigma_{i,ring}$ is the total diameter of the $i$-th ring, i.e. $\sigma_{i,ring}=2R_{i,H}+\sigma_{i,m}$, being $R_{i,H}$ and $\sigma_{i,m}$  respectively the radius of the $i$-th ring and { the diameter} of each of its monomers, due to the polydisperse ring/monomer distributions. 

\begin{figure*}[t!]
\includegraphics[width=1\textwidth]{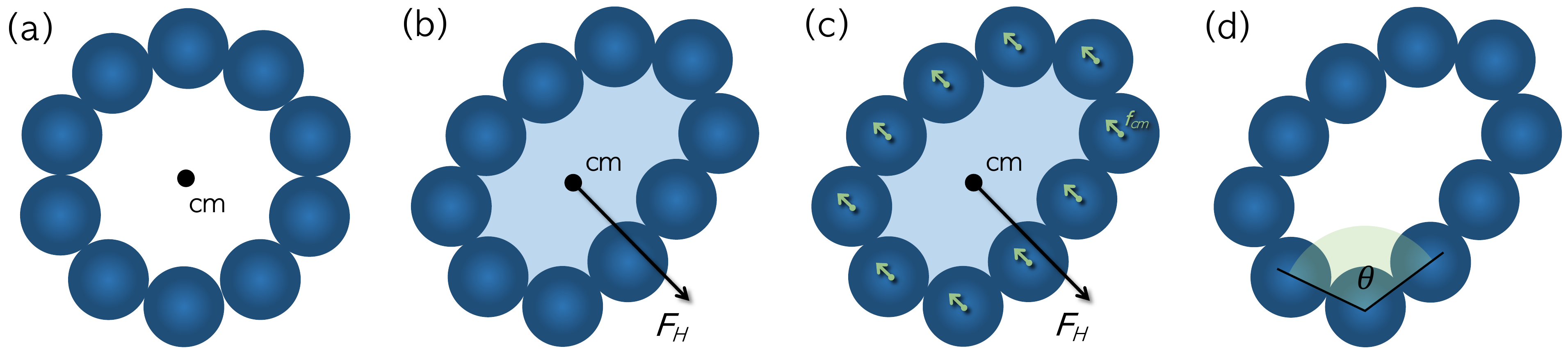}  
\caption{\label{fig:EPR} (a) Zero-stress polymer ring with no force acting on the center of mass as no deformation occurs; (b) Elastic Polymer Ring (EPR) deformed by thermal fluctuations or neighboring rings generating a net non-zero force $\vec{F}_H$ on its center of mass (cm); (c) equilibrium Elastic Polymer Ring (eq-EPR) where $\vec{f}_{cm}$ is re-distributed among all monomers; (d) semi-flexible polymer ring (SFPR) where an angular harmonic potential among particles introduces stiffness in the ring.}
\end{figure*}

To provide the rings with internal elasticity, different additional interactions are considered in each model.  In the EPR model introduced in Ref.\cite{Gnan},  monomers belonging to each ring also interact with an internal elastic repulsion that 
is modeled for simplicity by a hertzian field~\cite{Doghri} of strength $U$ that acts between each monomer and the center of mass of the ring. The hertzian field is defined as
\begin{equation}
V_H (r) = U\left(1-\frac{r}{R_H}\right)^{5/2}\Theta\left(1-\frac{r}{R_H}\right)
\label{eq:hertzian}
\end{equation}
\noindent { where $\Theta$ is the Heaviside step function}.
The effect of the field on the monomers belonging to the same ring is to maintain its circular shape; therefore the variation of the value of $U$ gives rise to a different ability of the particle to deform, and hence to a different softness.
The presence of the hertzian field originates a net resultant force acting on the ring center-of-mass $\vec{F}_{H}$ that is strictly zero only for purely symmetric rings, as shown in Fig.~\ref{fig:EPR}(a). However, the net force is non-zero for non-symmetric configurations (see Fig.~\ref{fig:EPR}(b)) as a consequence of thermal fluctuations and/or of the excluded volume interactions with neighboring particles. 

In the equilibrium version of the EPR model, i.e. the eq-EPR model, we balance the force  $\vec{F}_{H}$ acting on the center of mass by 
redistributing its opposite onto each monomer belonging to the same ring, i.e. $\vec{f}_{cm}=-\vec{F}_{H}/N_{m}$, as illustrated in  Fig.~\ref{fig:EPR}(c). 
{ Since $\vec{f}_{cm}$ depends on ring deformation, monomers of different rings are subjected to different $\vec{f}_{cm}$.  The presence of such a force inevitably influences inter-ring interactions, thus affecting ring deformation at high densities. 
As we show below, the net effect of this restoring force is to reduce the ability of the particles to deform at high packing fractions.}


Finally, we examine a third model of semi-flexible rings, where there is no hertzian field, but instead we consider an angular harmonic potential acting between three consecutive monomers, shown in Fig.~\ref{fig:EPR}(d). This reads as,
\begin{equation}
V_{\theta}(r) = k_{\theta} (\theta - \theta_0)^2
\label{eq:angle}
\end{equation}
where the equilibrium angle $\theta_0$ is chosen to be $\pi$, similarly to other models for semi-flexible polymers\cite{wu2018folding,englebienne2012directional,matysiak2006dynamics,huang2016simple},
while $k_{\theta}$ is varied to change the internal elasticity, i.e. the softness, 
of the particles.

For the three models, we perform Langevin dynamics simulations at constant temperature with {$k_BT/\epsilon=1$}. Length and time are given in units of  the average ring diameter $\langle\sigma_{ring}\rangle$ and of $t_0=\langle\sigma_{ring}\rangle \sqrt{m_{ring}/\epsilon}$, where $m_{ring}=m\cdot N_{m}$ and $m$ is the monomer mass which is set to unity. A velocity Verlet integrator is used to integrate the equations of motion with a time step $dt=10^{-3}$.  For EPR and eq-EPR simulations, we model Brownian diffusion following Ref.~\cite{Russo_2009} by defining the probability $p$ that a particle undergoes a random collision every $Y$ time-steps for each particle. By tuning $p$ it is possible to obtain the desired free {monomer} diffusion coefficient $D_0=(k_BTY dt/m)(1/p -1/2)$. { It can be shown that, in the low density limit, the free diffusion coefficient of a ring is $N_{m}$ times smaller than $D_{0}$.} Since by changing $D_0$ there is no influence on the long-time behavior, we fix $D_0=0.008$ for EPRs and $D_0=0.08$ for eq-EPRs. For the SFPR model, we perform Langevin dynamics simulations with $D_0=0.1$ using the LAMMPS simulation package\cite{plimpton1995fast}.


We investigate the static and dynamic properties as a function of $\zeta$ for EPRs with $U=100, 200, 500$ and  $1000$. We also study standard bead-spring rings, corresponding to the case $U=0$, for comparison.
In addition, we investigate the  eq-EPR model with  $U=30, 50, 80$ and $100$ and the SFPR models with $k_{\theta}=4, 5, 6, 7$. For all the models, parameters have been chosen in order to observe the onset of a reentrant transition~\cite{Gnan}. 
{ A detailed discussion of the choice of parameters is provided in the Supplementary Material (SM).}
For the dynamic features we study the mean-squared displacement (MSD) of the center of mass of each ring defined as $\langle \Delta r^{2}\rangle = \langle \overline{[\vec{r}_{cm}(t)-\vec{r}_{cm}(0)]^2}\rangle$ where $\overline{(\cdots)} = (1/N)\sum_{i=1}^{N}(\cdots)$ is the system average and $\langle \cdots\rangle$ is the time average. Moreover we investigate the evolution of the relaxation time $\tau_{\alpha}$ extracted from the self-intermediate scattering function $F_{s}(q^*,t)=\langle \overline{\exp{[\vec{q^*}\cdot (\vec{r}_{cm}(t)-\vec{r}_{cm}(0))]}}\rangle$ where $\vec{q^*}$ is the wave vector at which relevant interactions take place. $\tau_{\alpha}$ is defined as the time at which $Fs(q^*, t) = 1/e$ (where e is Euler’s number). Finally we study the temporal self-correlation function for a number of quantities defined as follows: given an observable $\mathcal{O}$ the correlation function employed  is defined as $C_{\mathcal{O}}(t)=\langle \overline{\frac{\mathcal{O}(t)\mathcal{O}(0)-\langle \mathcal{O}\rangle^2}{\mathcal{O}^2(t) -\langle \mathcal{O}\rangle^2}} \rangle $.
Regarding the static quantities, we show the distribution function and the fluctuations of the asphericity parameter defined as $a=[(\lambda_{2}-\lambda_{1})^2]/[(\lambda_{1}+\lambda_{2})^2$]~\cite{Rudnick1986}, where $\lambda_{1}$ and $\lambda_{2}$ are the eigenvalues of the gyration tensor of the ring.

Additionally, to gain more insight into the anomalous dynamics of EPR we introduce a modified, hertzian model, that consists of polydisperse disks with the same polydispersity as the EPRs undergoing Langevin dynamics  with $D_0=0.008$.  The disks interact both with the standard hertzian potential $V(r_{ij})=U_{H}(1-r_{ij}/\sigma_{ij})^{2.5}$ with $U_{H}=150$ and with an additional term resulting from a so-called overlap force acting on their centers of mass, defined as {$F^{i}_{A}(\vec{r}_{ij})=-K(A_{ov}^{ij}/A^{i})\cdot \vec{r}_{ij}$}, where $\vec{r}_{ij}$ is the vector distance between disk $i$ and $j$, $\sigma_{ij}=\frac{1}{2}(\sigma_{i}+\sigma_{j})$ is the average size of the two particles, $A^{i}$ is the area of disk $i$, $A_{ov}^{ij}$ is the overlap area between two disks $i$ and $j$, and $K$ is the amplitude of the force. In the case of monodisperse particles, the overlap force would be symmetric i.e. $F^{i}_{A}(\vec{r_{ij}}) = -F^{j}_{A}(\vec{r}_{ij})$. However, due to the polydispersity of the system, we have that $A^{i}\neq A^{j}$ which results in a non-symmetric force. This generates an out-of-equilibrium dynamics in analogy with the hertzian force that is present in the EPR system. The different nature of the two models is accounted by the way in which such a force is originated: through ring deformation for EPRs  and through particle overlaps for modified hertzian disks.




\section{Results}\label{sec:results}

\subsection{Effects of the hertzian field on the dynamics of EPR: equilibrium vs non-equilibrium behavior}\label{sec:eq_vs_noneq}
 \begin{figure}
 \centering{
    \includegraphics[width=.45\textwidth]{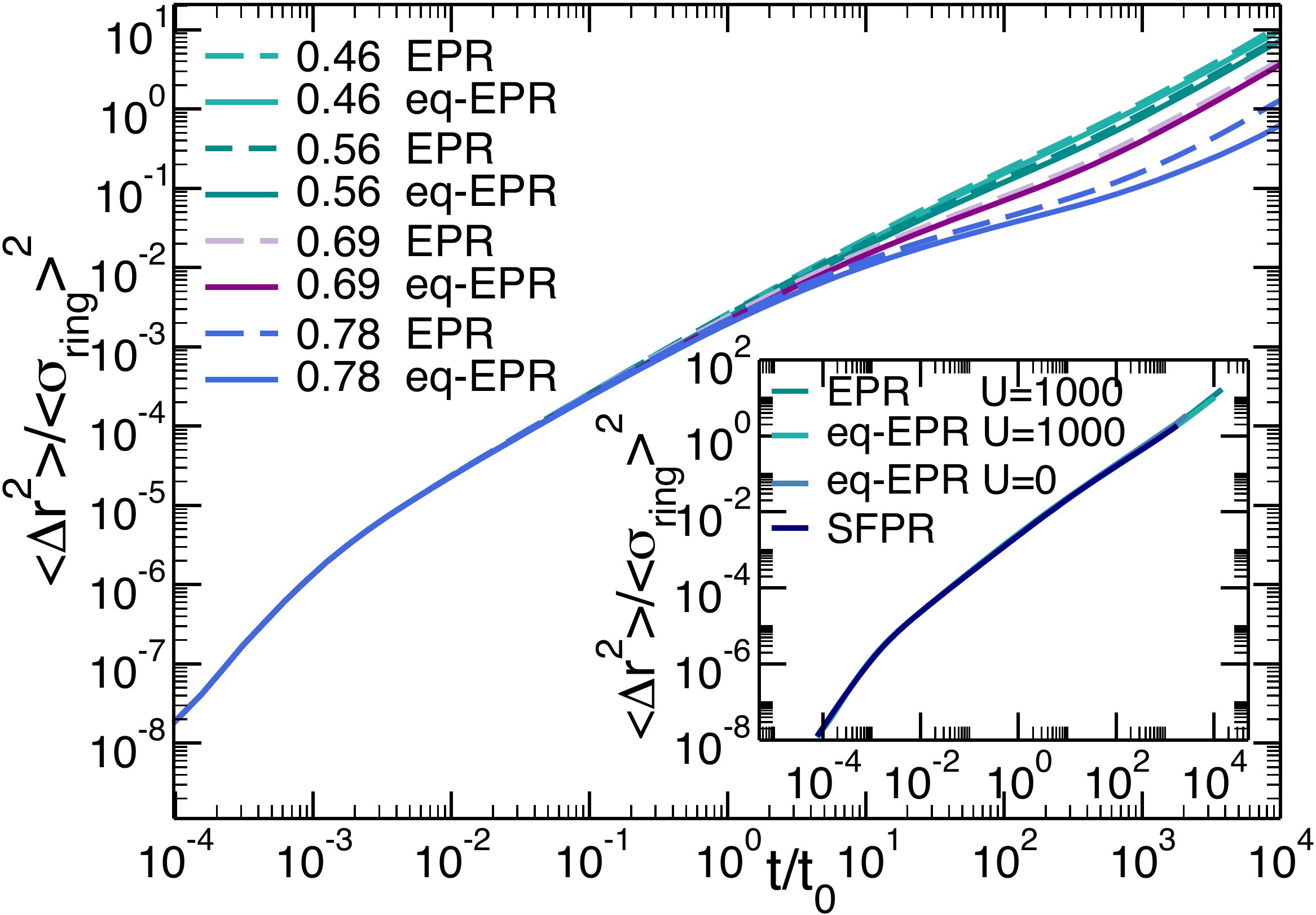}
    \caption{\label{fig:low_zeta} Mean-squared-displacement (MSD) of EPR and eq-EPR model  with $U=1000$ as a function of $\zeta$. Inset: MSD of different ring models at $\zeta=0.46$. Here, $D_0=0.008$ for all models. }
    }
\end{figure}
\begin{figure*}[ht!]
\includegraphics[width=1.\textwidth]{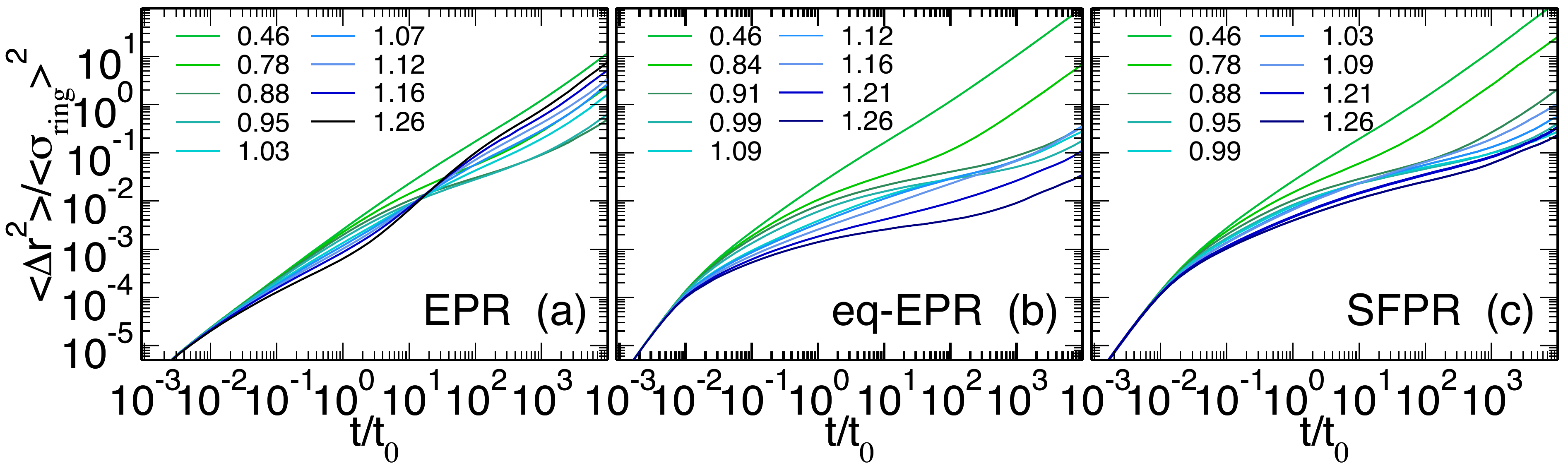}
\caption{\label{fig:msd} 
Mean-squared displacement as a function of reduced time $t/t_0$ at different packing fractions $\zeta$ for the three different models: (a) EPRs with $U=200$, (b) eq-EPRs with $U=100$ and (c) SFPRs with $k_{\theta}=5$.} 
\end{figure*}
\begin{figure*}[!ht]
\includegraphics[width=1.\textwidth]{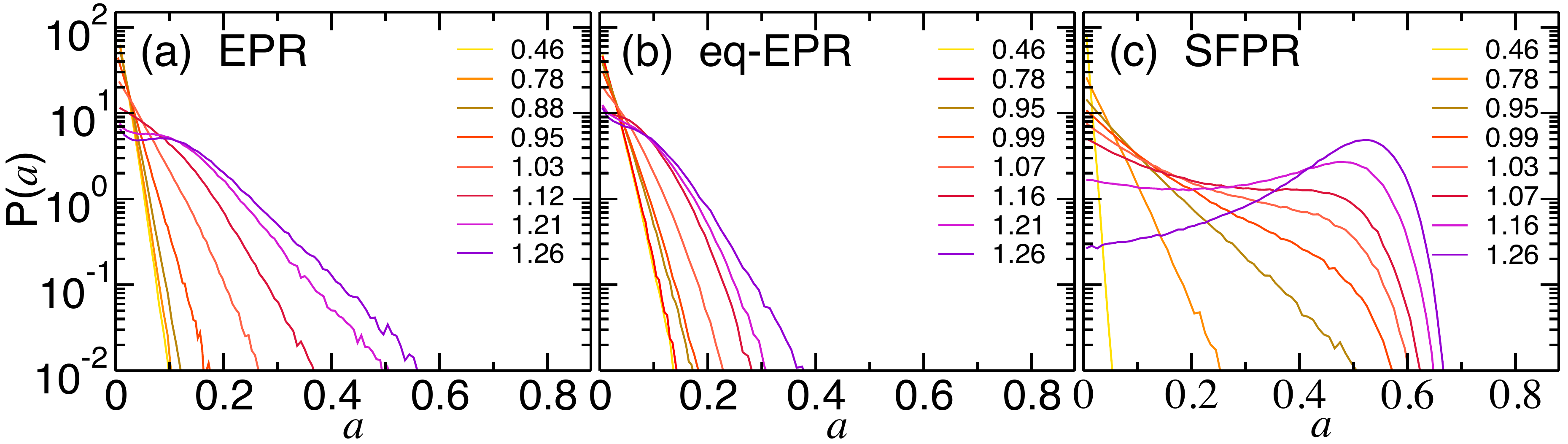}
\caption{\label{fig:asphdistrib} 
Asphericity distribution for the three different models at different packing fractions $\zeta$: (a) EPRs with $U=200$, (b) eq-EPRs with $U=100$ and (c) SFPRs with $k_{\theta}=5$.}
\end{figure*}

\begin{figure*}[!ht]
\includegraphics[width=1.0\textwidth]{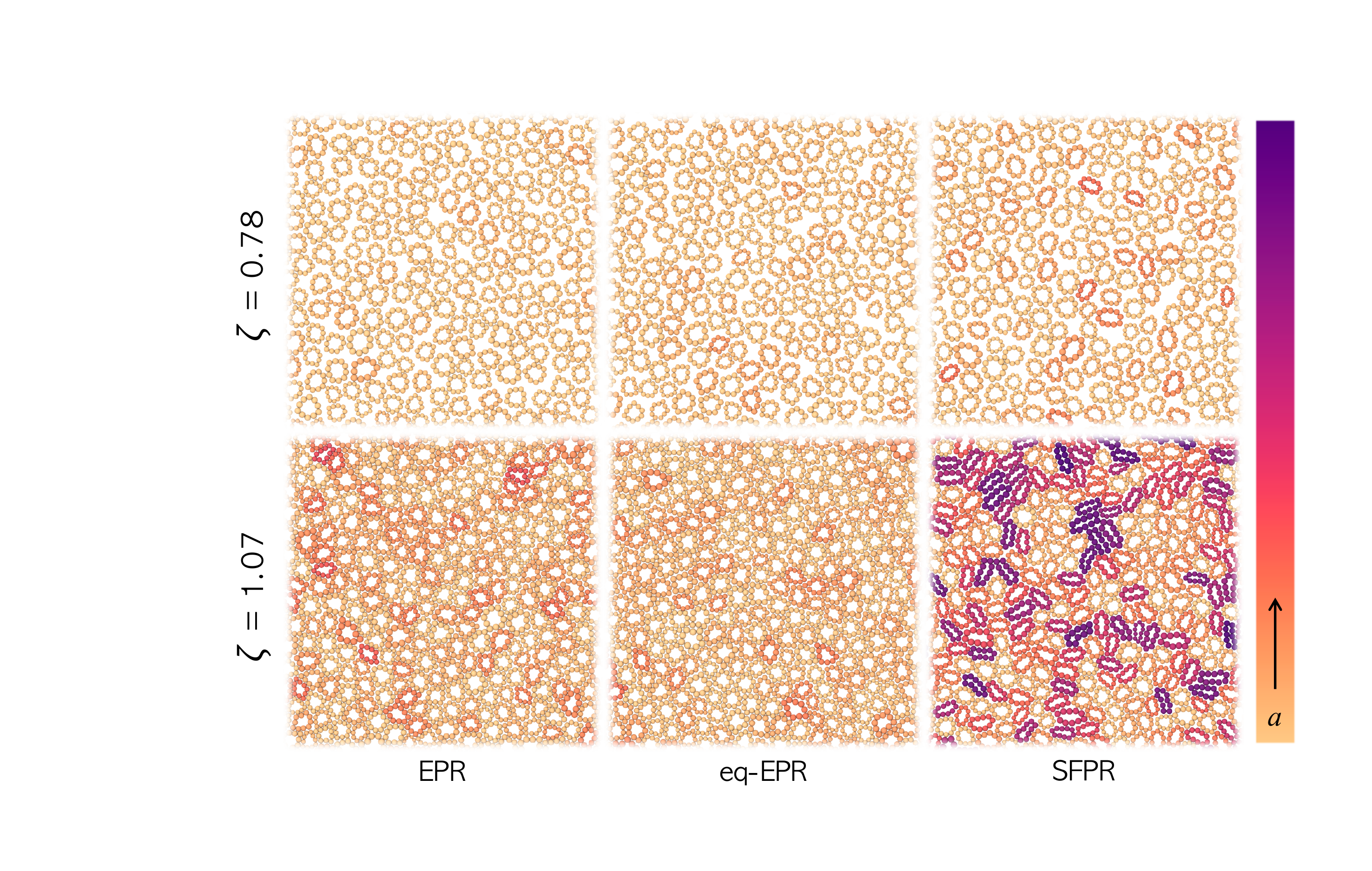}
\caption{\label{fig:snaps} 
Simulation snapshots displaying the three models analyzed at two representative packing fractions, namely $\zeta=0.78$ and $1.07$: for EPRs $U=200$, eq-EPRs $U=100$ and SFPRs $k_{\theta}=5$. Color-coding reflects the asphericity of each ring in the different models.}
\end{figure*}

In this section, we analyze the influence of the out-of-equilibrium features generated by the hertzian force in the EPR model by comparing its dynamical behavior and its ability to deform with that of the eq-EPR and of the SFPR model.
The presence of the hertzian field in the EPR model generates an unbalanced internal force $\vec{F}_{H}$ that acts on the center of mass whenever the ring is deformed. We can consider $\vec{F}_{H}$ in all respects as an active force that, contrarily to standard active systems~\cite{Fily:2012aa, Maggi_2015}, depends on the degrees of freedom of the deformed ring.
Ring deformation can occur for two main reasons: due to thermal effects, or due to the mechanical action of neighboring particles that via the excluded volume interactions (i.e. WCA interactions) deform the ring. The first scenario dominates at low $\zeta$, while the second one occurs at high $\zeta$, beyond the jamming volume fraction, when rings can fill the available space only by deforming.  
In the dilute regime,
the tiny deformations induced by thermal fluctuations are so small that the generated force acting on a single monomer does not  effectively contribute to the dynamics as compared to the deterministic forces (originated by FENE and WCA contributions), so that the system behaves as in equilibrium. This is illustrated in Fig.~\ref{fig:low_zeta} (inset), where we compare the MSD  of the EPR model for different type of rings and values of the elastic strength, including $U=0$, i.e. in the absence of the hertzian field at low packing fraction ($\zeta=0.46)$. We also report in the main figure the comparison between EPRs and eq-EPRs at the same $U$ as a function of $\zeta$ but still below the  jamming point. On increasing $\zeta$, rings collisions increase, causing larger shape fluctuations in the EPRs which slightly enhance their diffusion with respect to the eq-EPRs. As a consequence, $\vec{F}_{H}$ starts to play a role in the dynamics.
However it is at high $\zeta$ that excluded volume interactions which continuously deform EPRs give rise to large contributions of $\vec{F}_{H}$.
To clarify this point we now compare the dynamics of the EPR system with that of different models of polymer rings for selected values of softness (U=200 for EPRs, U=100 for eq-EPRs and $k_{\theta}=5$ for SFPRs). { Such a selection is based on the fact that the fragility parameter, that will be introduced later  on in the text, is roughly the same for the three models.}

 Figure~\ref{fig:msd} shows the MSDs for EPRs, eq-EPRs and SFPRs, respectively, at different $\zeta$ values from low densities up to and above close contact.  The first important finding of our analysis is that  a reentrant dynamics, albeit much less pronounced, takes place also for eq-EPRs and SFPRs, as for the EPR model. This implies that the system initially gets slower with increasing packing fraction, and then it speeds up again roughly above $\zeta \sim 0.9$, which signals the packing fraction where the rings are still largely undeformed and in close contact with each other (a loose jamming definition){ \footnote{\label{note0} For monodisperse hard-disks the maximum close packing is set to $\zeta=0.82$. This can be considered as a lower bound to the actual jamming volume fraction for polydisperse disks.}}. Above this value, the internal degrees of freedom of the rings start to play an important role and, through shrinking and deformation, the dynamics gets faster. However, in contrast to the EPR model, such a reentrance occurs in a limited $\zeta$ window for eq-EPR and SFPR models because, at large enough $\zeta$, a new slowing down mechanism takes place, finally leading to an arrested state for both models.  Such arrest is instead not found for EPRs in the whole investigated $\zeta$-region.
In addition, while the super-diffusive regime is clearly observed in the EPR system at large $\zeta$, no sign of super-diffusion is present in the MSDs of the other two systems, independently of the employed model parameters.

With the aim of identifying the differences among the three models, we also compare them in terms of particle asphericity $a$, in order to understand how different types of rings respond to large mechanical compressions. Fig.~\ref{fig:asphdistrib} reports the asphericity distribution $P(a)$ for the three models and different packing fractions. We first notice that, as expected, eq-EPRs have a much reduced tendency to deform { at high $\zeta$} as compared to EPRs. 
On the other hand, when looking at the asphericity distribution of the SFPRs, we notice that rings are able to achieve very large asphericity values at high $\zeta$. This leads to the emergence of a double-peak distribution at low and high asphericity, respectively, which  indicates that there are two populations of particles, one which is almost undeformed and the other that is highly deformed.
Representative snapshots of the three systems at low and high $\zeta$ are shown in Fig.~\ref{fig:snaps}, where particles are colored according to their asphericity, clarifying the differences between the models. While at $\zeta =0.78$ the three systems look rather similar, we find that at $\zeta=1.07$, the SFPR model displays many more rings that are largely deformed with respect to the EPR system. 
\begin{figure*}[!htb]
\includegraphics[width=1\textwidth]{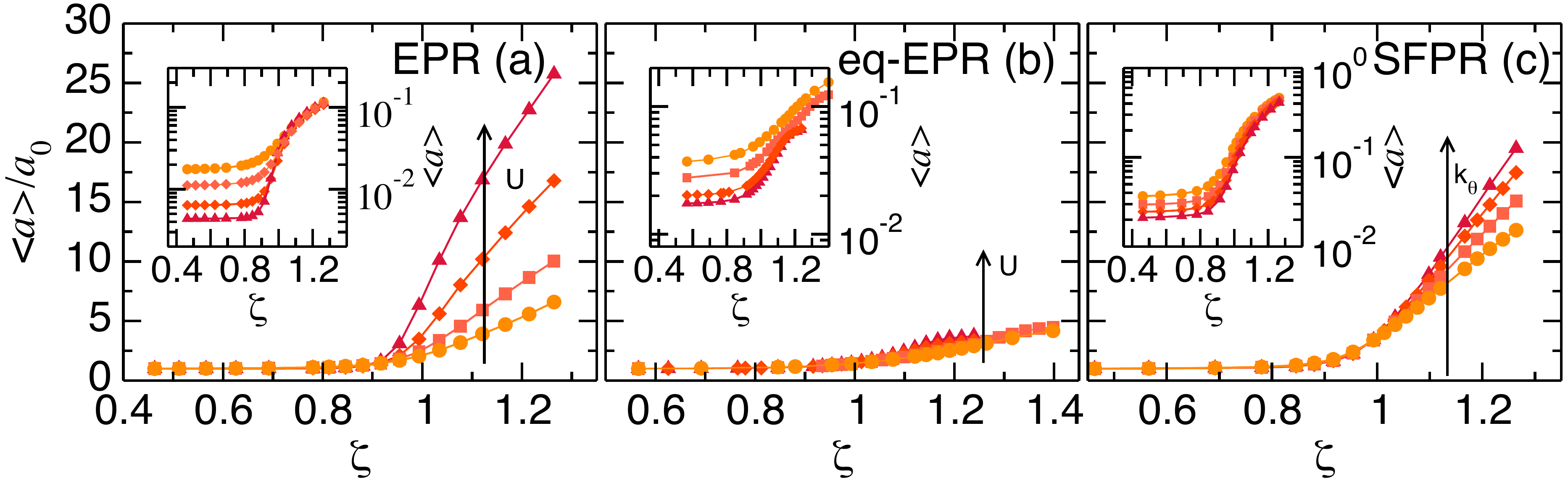}
\caption{\label{fig:asph} 
Average asphericity normalized by its low-$\zeta$ average value,  $\langle a\rangle/a_0$, as a function of the packing fraction $\zeta$ for: (a) EPRs with $U=100, 200, 500, 1000$ and $D_0=0.008$, (b) eq-EPRs with $U=30, 50, 80, 100$ and $D_0=0.08$, (c) SFPRs with $k_{\theta}=4, 5, 6, 7$ and $D_0=0.08$. Insets: same data without the low-density normalization.}
\end{figure*}
 \begin{figure*}[!ht]
\includegraphics[width=1\textwidth]{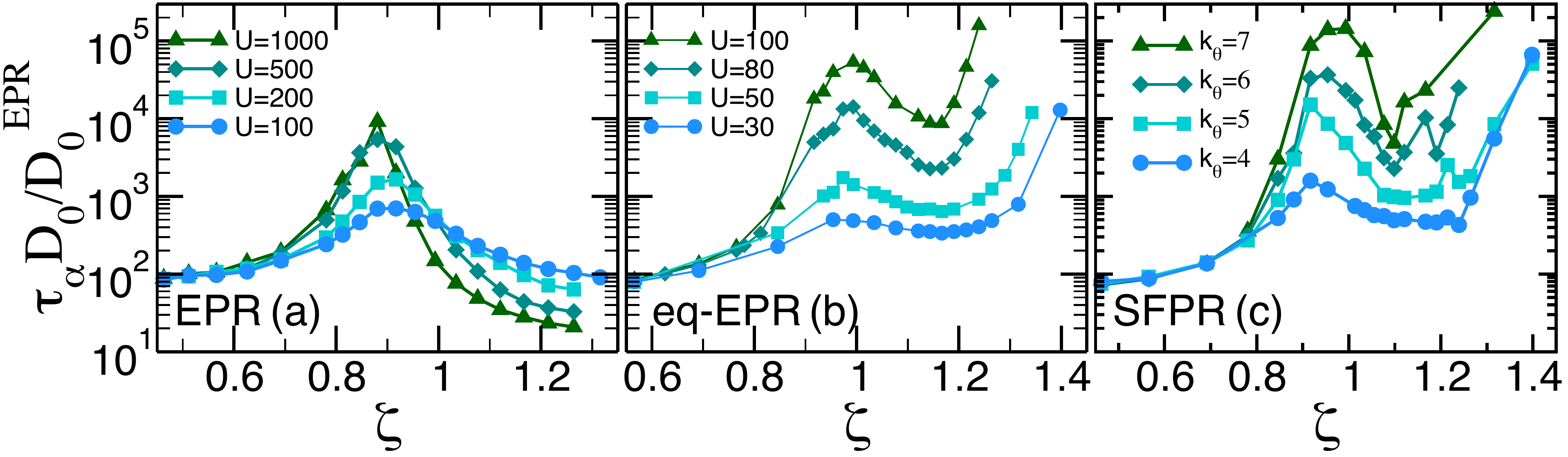}
\caption{\label{fig:tau_rings} 
Rescaled relaxation time as a function of the packing fraction $\zeta$ for: (a) EPRs with $U=100, 200, 500, 1000$ and $D_0=D_0^{EPR}=0.008$, (b) eq-EPRs with $U=30, 50, 80, 100$ and $D_0=0.08$,(c) SFPRs with $k_{\theta}=4, 5, 6, 7$ and $D_0=0.1$.}
\end{figure*}
\noindent  Fig.~\ref{fig:asph} shows the normalized average asphericity for the three models at several $\zeta$ for different parameters. The normalization consists in dividing $\langle a\rangle$ by  $a_{0}$, that is the low-density value of the average asphericity.  There is a striking difference between the behavior of eq-EPRs in Fig.~\ref{fig:asph}(b) as compared to EPRs and SFPRs, respectively, in Fig.~\ref{fig:asph}(a) and (c). Indeed, the poor ability of the equilibrium rings to deform strongly limits the change in asphericity, not only upon increasing $\zeta$, but also on changing $U$.

\subsection{Link between deformation and fragility}\label{sec:fragile}
 We now discuss more in detail the reentrant dynamics mentioned in the previous paragraph for the three models. To better understand the extent of such phenomenon for the different ring  models, we have performed simulations at several $\zeta$ values for different parameters of the eq-EPR and SFPR models to compare them with the EPR system. For each state point we have calculated the self-intermediate scattering function $F_{s}(q^*,t)$ from which we have extracted the relaxation time $\tau_{\alpha}$ (in reduced units). The relaxation times for the three models of polymer rings and varying softness, as a function of $\zeta$, are shown in Fig.~\ref{fig:tau_rings}. { $\tau_{\alpha}$ has been rescaled for $D_0/D_{0}^{EPR}$ in order to have comparable relaxation times for the three models  at low packing fractions.}
\noindent For all cases, we observe an initial increase of $\tau_{\alpha}$ from low $\zeta$ up to the jamming point, followed by a sudden decrease that signals a speed-up of the dynamics, whose variation depends on the model of ring employed and on its softness. The main difference between the EPRs and the other two models is that in eq-EPRs and SFPRs the reentrance is limited to a finite region  of packing fractions whose width again depends on the specific system and the employed softness parameter. Hence, despite both eq-EPRs and SFPRs display a reentrant transition, at higher packing fractions the relaxation time increases again, signalling the onset of the dynamical arrest. 

\begin{figure}[!h]
\includegraphics[width=1.\linewidth]{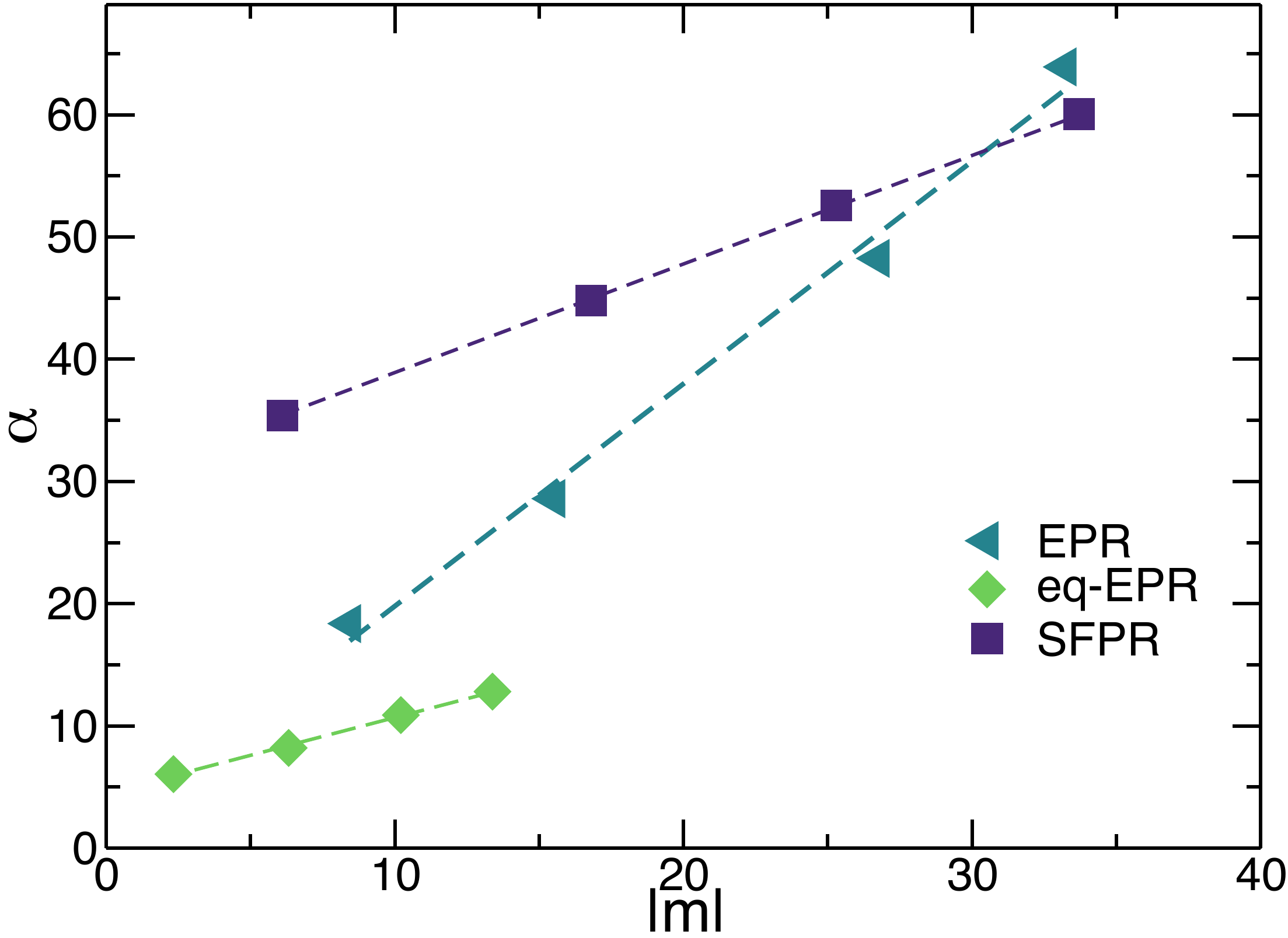}
\caption{\label{fig:fragility_plot} Asphericity variation $\alpha$ as a function of the absolute value of the fragility $m$ for the three models of polymer rings. Dashed lines are linear fit used as guides to the eye.}
\end{figure}

These results show that, for all the studied ring models, the dynamics speeds-up because of ring deformation. We can thus ultimately answer the question on whether the dependence of the dynamics on softness is a general feature of elastic particles, as hypothesized in Ref.~\cite{Mattsson2009}, or it is just a peculiar property of the EPRs. In Ref.~\cite{Gnan} some of us have addressed this problem by studying the relation between fragility\footnote{\label{note2} For the EPR system, the $\zeta^*$ value was chosen to be close to the packing fraction  $\zeta_{R}$ at which the reentrance occurs with $\zeta^{*}>\zeta_{R}$} defined as,
\begin{equation}
m=[d(\ln\tau_{\alpha})/d(\zeta/\zeta^{*})]|_{\zeta=\zeta^{*}},
\end{equation}
and deformation described in terms of the asphericity variation:
\begin{equation}
\alpha=(1/a_{0})[d\langle a\rangle/d\zeta]|_{\zeta=\zeta^{*}}.
\end{equation}
In Both definitions, $\zeta^*$ is the (U-dependent) value of the packing fraction at which $\tau_{\alpha}$ is the same for all U just after the reentrant regime.
For the eq-EPR model,  it is immediately evident from Fig.~\ref{fig:tau_rings}(b) that it is not possible to find a common time $\tau_{\alpha}$ among the different curves since they are well separated for $\zeta>\zeta_{R}$. For the SFPR system in Fig.~\ref{fig:tau_rings} (c) a common time for some of the curves can still be found, but this would allow us to determine the fragility for only two values of $k_\theta$. In order to circumvent this problem, we employ an alternative procedure where we calculate the fragility from the behavior of the relaxation time for $\zeta^{*}\equiv \zeta_{R}$. By applying this strategy also in the analysis of the EPRs, we find that the newly calculated values of $m$ and $\alpha$ do not change significantly as compared to previous findings in Ref.~\cite{Gnan}.
Analogously to Fig.~\ref{fig:asph} we extract $\alpha$ with a linear fit of the curves for $\zeta>\zeta_{R}$. Our aim is to verify that a linear relation exists between $m$ and $\alpha$, independently of the model. The results are shown in Fig.~\ref{fig:fragility_plot}, where it is evident that, for all investigated models, a linear relation between fragility and elasticity is always valid. Thus, this is not just a feature of the EPR model. 
Interestingly, the different slopes of the linear relations between $|m|$ and $\alpha$ highlight the different ways in which the rings are able to respond both in terms of deformation and of dynamical properties to a change in $\zeta$. In particular, considering Fig.~\ref{fig:asph}, we expect eq-EPRs to be those with the smaller $\alpha$ vs $|m|$ slope due to the minimal change in $\alpha$ upon varying the system parameters. This is indeed consistent with the findings in Fig.~\ref{fig:fragility_plot}.

\subsection{Anomalous dynamics}
In Sec.~\ref{sec:eq_vs_noneq}  we have shown that SFPRs are able to deform even more than EPRs, while the asphericity of eq-EPRs is small if compared to the other two models. The ability to deform is a necessary, but not sufficient condition for the uptake and release of the stress which has been proven to be the key ingredient for the observation of an anomalous dynamics. In this context, several  experimental studies, mainly on colloidal gels,  have shown that the dynamics is sometimes faster than exponential~\cite{cipelletti2000universal, Augusto_de_Melo_Marques_2015, Dallari2020, Nigro_2020, Pastore2020, Jain2020}, i.e. it was observed that  the intermediate scattering function $F(\vec{q},t)$ at a typical wavevector $\vec{q}$ can be described by a generalized exponential decay $F(\vec{q},t)\sim\exp(t/\tau)^{\beta}$ (where $\tau$ is the relaxation time) with an exponent $\beta$ greater than 1.0. At the microscopic level, $\beta>1.0$  implies that particles move faster compared to standard diffusion, i.e. the motion is super-diffusive at the investigated length scale. These experimental evidences led to the formulation of some hypothesis on the microscopic mechanism that generates the occurrence of a super-diffusive behavior and which involves stress propagation in colloidal gels. In particular, it has been argued that in these systems the reorganization of the network follows some micro-collapses, in which bonds among particles are broken thus releasing stress into the network and triggering a super-diffusive motion of neighboring particles~\cite{cipelletti2000universal}. The possibility to achieve a faster than exponential dynamics has been established  in mean-field models of elastic materials where the disruption of the network has been modeled as a number of Poissonian events  that act  as  dipole forces with long-range elastic effects~\cite{Pitard2002}. More recently, numerical simulations have shown that a compressed exponential decay of the density correlators is possible if single bonds are selected and artificially broken in order to observe stress propagation within local environment~\cite{bouzid2017elastically}. Hence, it follows that stress propagation seems to be the key ingredient that is needed to observe a super-diffusive dynamics. For the EPR particles, super-diffusion is observed at intermediate time scales in the mean-squared displacement, accompanied by a compressed exponential relaxation in the self-intermediate scattering function, as a result of the superposition of regions which are dynamically heterogenous, among which clusters of particles that move ballistically~\cite{Gnan}. It has thus been argued that the microscopic mechanism responsible for this anomalous dynamics is the ability of the rings to deform that allows the release and the accumulation of the stress, triggering the super-diffusive dynamics at intermediate time scales. More than ring deformation, what should matter is the ability of the ring to vary its shape, i.e. to exhibit shape fluctuations. Hence, even if SFPRs deform more than EPRs, it is interesting to understand whether such very deformed rings are able to change their shape for the uptake and release  of the stress or, instead, whether they always remain deformed.  It is then important to consider the fluctuations of the asphericity in order to assess whether a ring is able to release the stress accumulated trough deformation within a certain time. 

\begin{figure}[!h]
\includegraphics[width=1.0\linewidth]{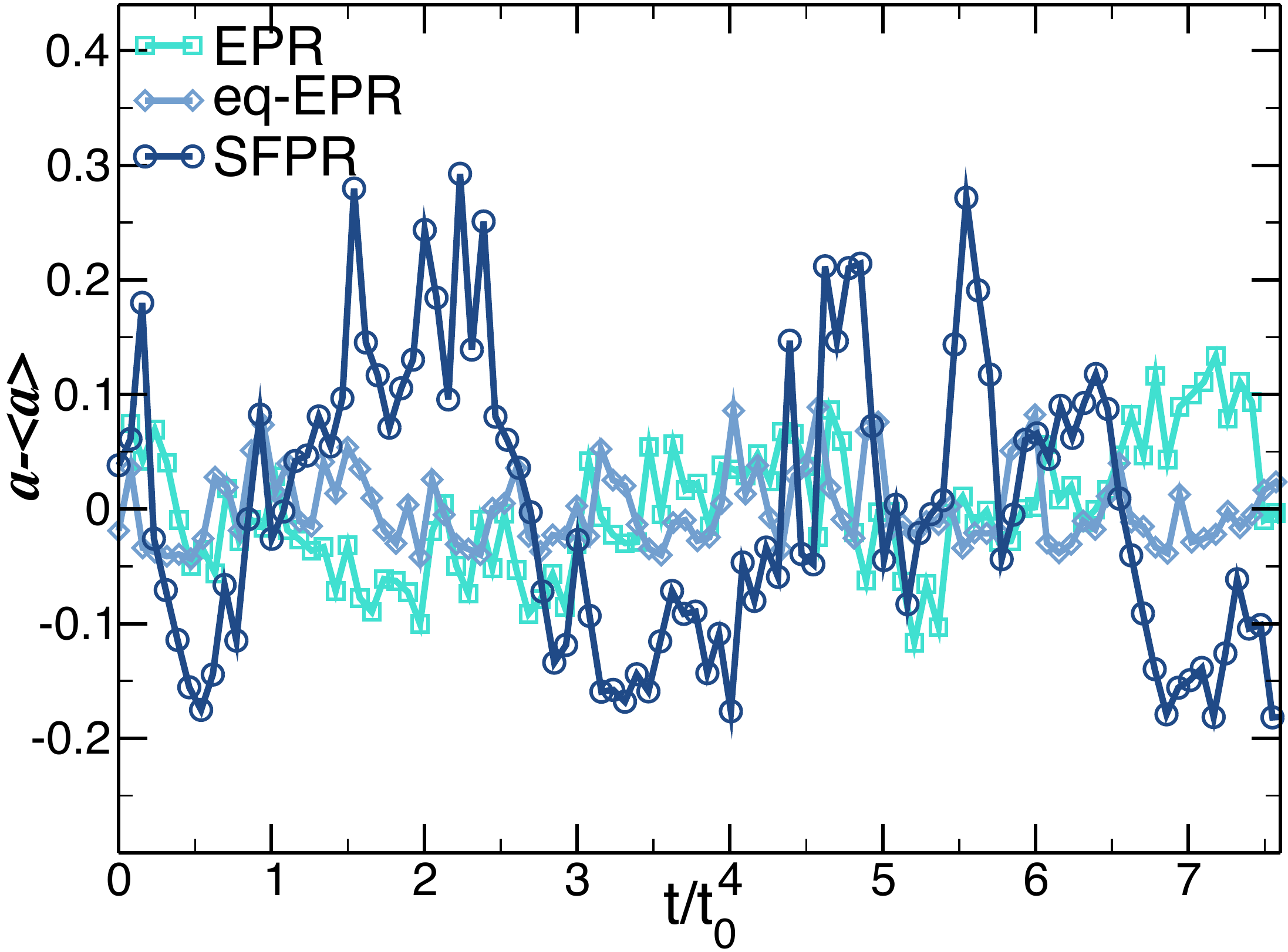}
\caption{\label{fig:AsphFluctuations} 
Fluctuations of the asphericity $(a-\langle a \rangle)$ for a single ring in the three models in the reentrant region: EPRs with  $U=200$ and $\zeta=1.26$, eq-EPR with $U=100$ and $\zeta=1.09$, SFPR with $k_{\theta}=5$ and $\zeta=1.14$.
For the three models, the monomer free diffusion coefficient was set to $D_0=0.08$.}
\end{figure}

To this aim, we compare the time evolution of the fluctuations of the asphericity of a single, representative ring for the three models in Fig. ~\ref{fig:AsphFluctuations}.
\noindent We find that the SFPR model fluctuates much more than EPRs and eq-EPRs, whose fluctuations are instead comparable. 
Thus, the semi-flexible rings not only display the most extreme deformations among the three investigated systems but also their asphericity fluctuations are the largest, which implies that their ability to release the stress is greater as compared to the EPR system. However, no sign of super-diffusion is detected in the SFPR model.

\begin{figure}[!h]
\includegraphics[width=1.\linewidth]{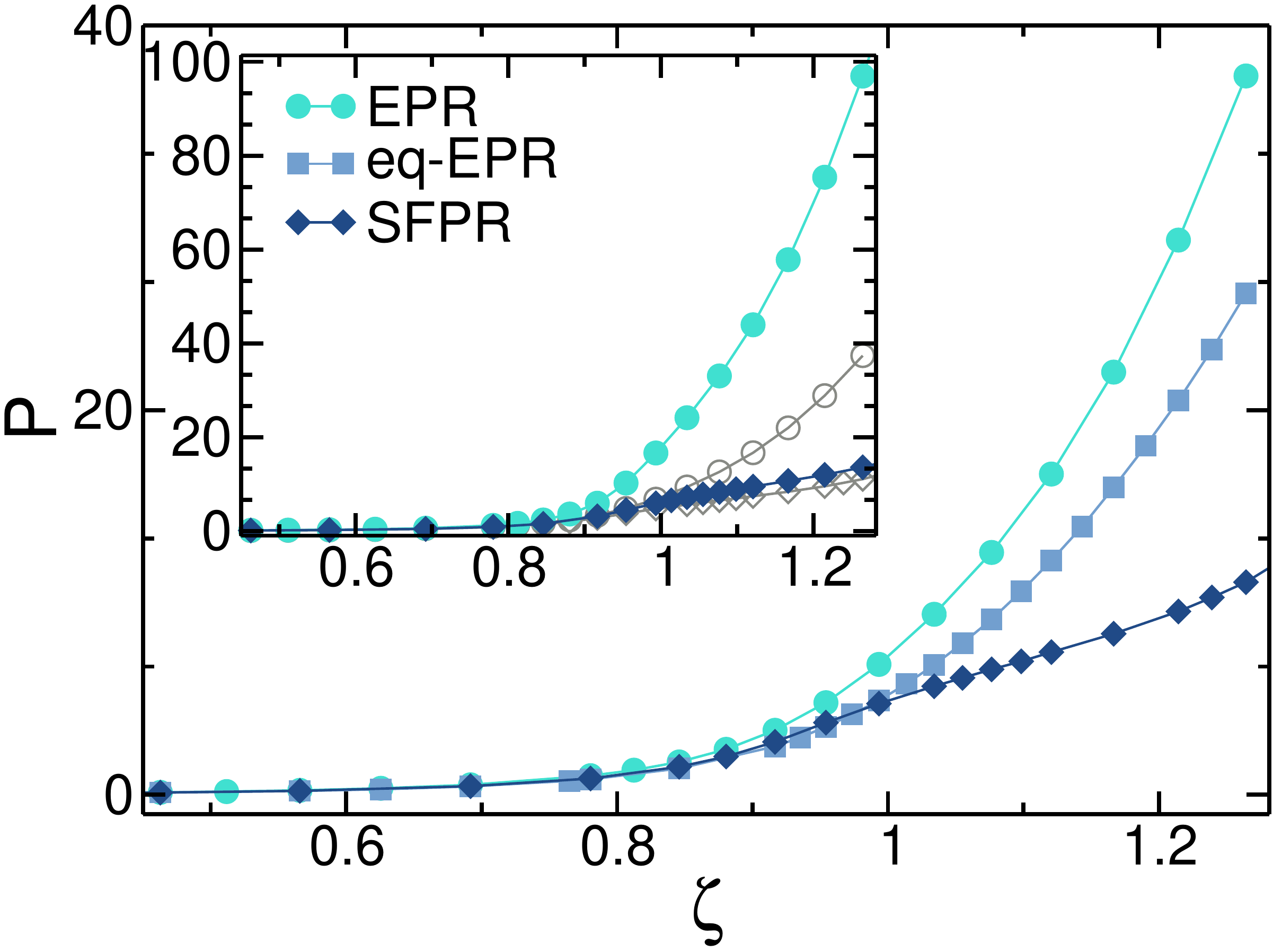}
\caption{\label{fig:Pressure} 
{ Main Panel: pressure $P$ as a function of $\zeta$ for EPRs with $U=200$, eq-EPRs with $U=100$ and SFPRs with $k_{\theta}=5$. Inset: pressure $P$ as a function of $\zeta$ for EPRs with $U=1000$ and SFPRs with $k_{\theta}=7$. As for the systems in the main panel, the parameters of  EPRs and the SFPRs have been chosen to provide similar fragility values. Gray curves are pressures for EPRs with $U=200$ and SFPR swith $k_{\theta}=5$ which are displayed for comparison.}}
\end{figure}
\noindent 
A hint to better understand this result comes from the comparison of the pressure $P$ of the three models, shown in Fig.~\ref{fig:Pressure}, where we observe that, for large $\zeta$, the pressure in the SFPR model is { always smaller than that of the EPR and eq-EPR systems. Importantly, the difference in pressure increases if different values of the parameters are chosen both for the EPR and the SFPR (inset of  Fig.~\ref{fig:Pressure}).} 
This implies that, despite larger fluctuations, { the amount of stress} that is { relaxed} by the SFPR model is overall smaller than for the other two. This feature, combined with the greater ability of  SFPRs to deform and fluctuate in shape, is probably enough to allow each ring to { relax} the stress by itself, without triggering collective phenomena and the associated super-diffusion. On the contrary, in the EPR model there is a force that continuously pumps energy into the system giving rise to a much higher total pressure. Since for EPRs shape deformation is large but shape fluctuations are small compared to SFPRs, we are left to speculate that the EPR system needs an extra mechanism { that allows an effective way to propagate stress among neighbor rings, that eventually results into the observed anomalous dynamics. }Regarding the eq-EPR system, there is small ring deformation and no stress { relaxation}, so it is legitimate to expect the absence of anomalous dynamics in this case.

\begin{figure}
 \centering{
    \includegraphics[width=.48\textwidth]{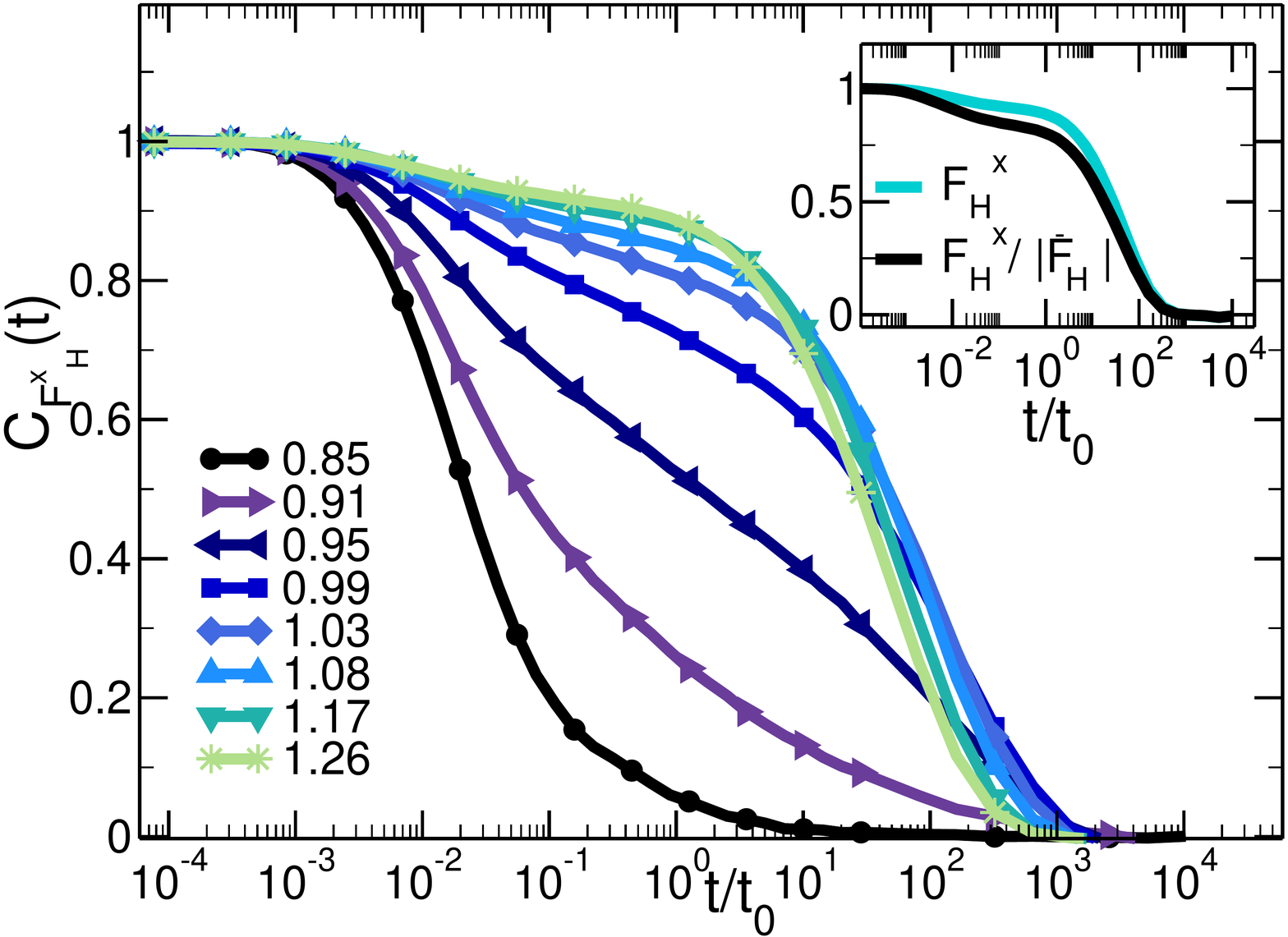}
    \caption{\label{fig:CorrActiveForce} Autocorrelation of the $x$-component of the hertzian force  $C_{F^x_{H}}(t)$ at different packing fractions $\zeta$ for EPRs with $U=1000$. Inset: Same as in the main panel for $\zeta=1.26$ compared to the correlation function of the $x$-component of  $\hat{F}_H=\vec{F}_H/|\vec{F}_H|$,  i.e. unit vector of the hertzian force.}}
\end{figure}

From the above considerations, we learnt that stress { relaxation} alone is not sufficient to trigger the occurrence of super-diffusion. Therefore, it may be a matter { either of total amount of stress in the system or of the effective propagation of this stress, that leads to the onset} of a collective effect as the one observed in Ref.~\cite{Gnan}. 
{ While we cannot definitely exclude the former hypothesis, given that the stress intensity for SFPR is always significantly smaller than for EPR, even for high values of $k_{\theta}$ (inset of Fig.~\ref{fig:Pressure}), we can examine in more detail the latter aspect. Indeed, we know that collective motion is often associated to active, non-equilibrium systems, seeming to imply that the presence of the unbalanced force in the center of mass of the rings could be the only cause of the occurrence of this additional mechanism leading to the anomalous dynamics.}
 For instance, super-diffusion can occur if there exists a persistent force that drives collectively particles towards the same direction. To quantify the persistence of $\vec{F}_{H}$, we evaluate the self-correlation function $C_{F_{H}}(t)$ of such force at different packing fractions. The results for the $x$-component of $\vec{F}_{H}$ are shown in Fig.~\ref{fig:CorrActiveForce}, confirming that for $\zeta \lesssim 0.85$, i.e. above the loosely-defined jamming point, the correlation function of the hertzian force quickly decreases
while, for larger values of $\zeta$, $C_{F_{H}}(t)$ displays a two step decay, with a plateau that extends over several decades in time and whose height increases on increasing $\zeta$. In addition, we find that the relaxation time of  $C_{F_{H}}(t)$ has a little dependence on $\zeta$ and roughly coincides with the time regime in which the system is characterized by long-time diffusion (i.e. $t/t_0 \gtrsim 10^2$ to be compared with MSDs shown in Fig.~\ref{fig:msd}(a)).
Fig.~\ref{fig:CorrActiveForce}  shows that a persistent force builds-up for each ring pointing towards a given direction for several decades (the plateau length) and then slowly decorrelates. The decorrelation of $\vec{F}_{H}$ could  result either from a change in the intensity or from a reorientation of the force. To disentangle the two contributions, we also evaluate the self-correlation of the $x$-component of $\hat{F}_H$, i.e. the unit vector of the hertzian force, which gives information on the orientation of the vector force. The inset in Fig.~\ref{fig:CorrActiveForce} shows that the latter  correlation function is quite similar to $C_{F_{H}}(t)$, meaning that most of the decorrelation occurs thanks to a change in the orientation of $\vec{F}_{H}$. 

Therefore, if the out-of-equilibrium force were the only responsible for the anomalous dynamics, then systems with a similar behavior to the EPRs could dissipate the stress with the same mechanism and display super-diffusive behavior. Here we show that this is not necessarily true by examining a simple soft system, very similar in spirit to the EPRs, which also self-generates an out-of-equilibrium force at high density, even though lacking the internal polymeric degrees of the rings. To this aim, we employ the modified hertzian model introduced in the Methods section, for which we study the dynamics at different packing fractions in order to identify whether the presence of the extra force gives rise to super-diffusion. We stress that the standard hertzian disks with the interaction strength employed here and in the absence of the extra force generated by the overlaps, already display a reentrant behavior at high $\zeta$, as reported in our previous work~\cite{Gnan}.

\begin{figure}[!h]
\includegraphics[width=1.0\linewidth]{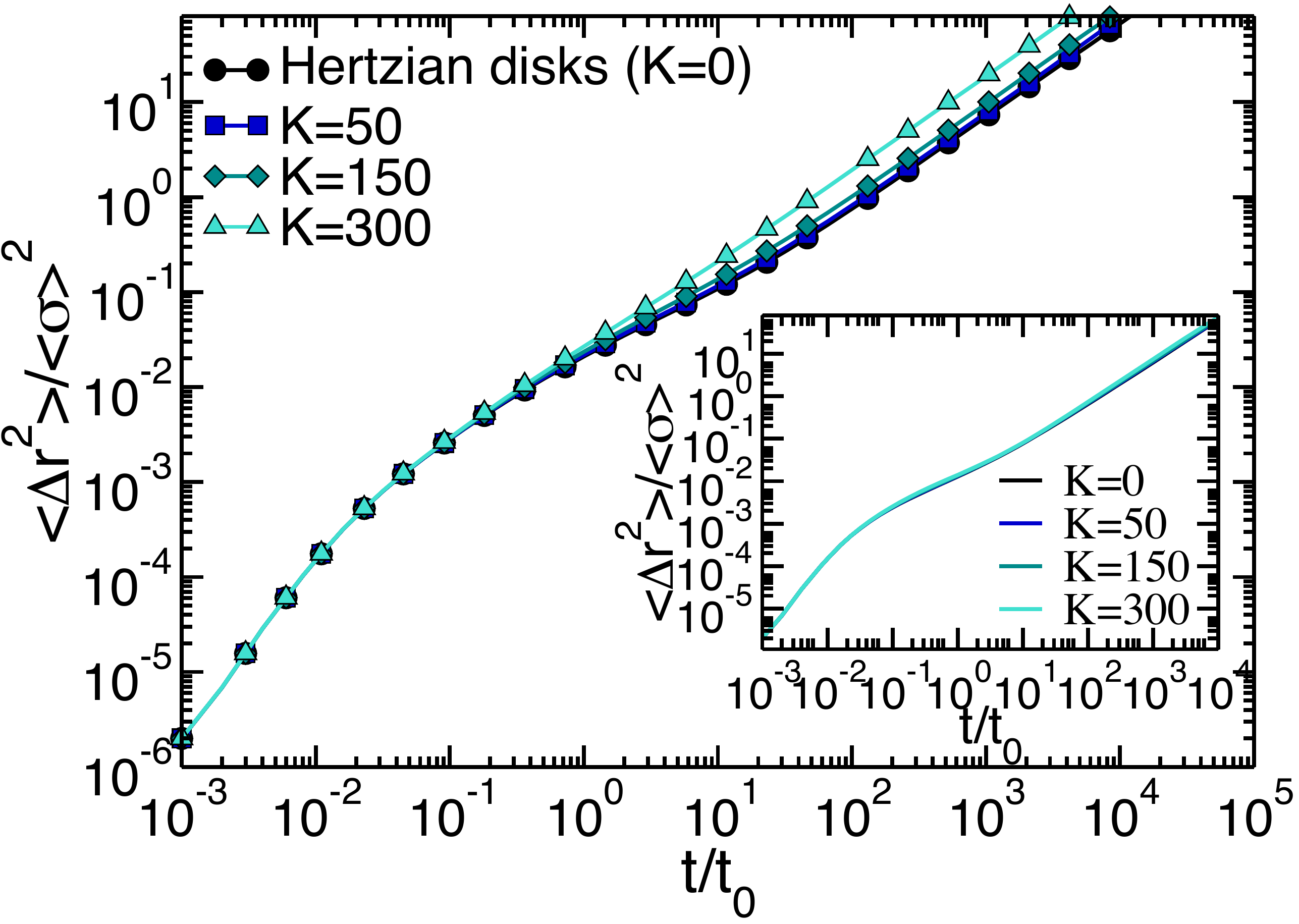}
\caption{\label{fig:ActiveDynamics} 
Main Panel: MSD for modified hertzian disks with $U_{H}=150$ at $\zeta=0.83$ as a function of the overlap force with different $K$. Inset: MSD as in the main panel but for $\zeta=2.53$ (reentrant point).}
\end{figure}
\noindent Figure~\ref{fig:ActiveDynamics} shows the MSD of the modified hertzian disks at high packing fraction upon increasing the amplitude $K$ of the overlap force: we see that the system becomes slightly faster for larger values of $K$. This is due to the fact that the force generated on disk $i$ due to the overlap with disk $j$ points towards $j$ (and viceversa for the force acting on $i$), which pushes disks to stay closer on increasing $K$, thus leaving more available space for rearranging and diffusing. This is also what happens in the EPR model when two rings $i$ and $j$ deform pushing one against the other: the resulting hertzian force that acts on the center of mass of ring $i$ points with good approximation towards the center of $j$  and viceversa. However, despite the analogies between the two systems,  for the modified hertzian disks the dynamics always remains diffusive, even if the auto-correlation force of the modified hertzian model also persists up to long times in analogy to the EPR model, as shown in the SM.
This behavior holds at all investigated $\zeta$, even above the reentrant point as shown in the inset of Figure.~\ref{fig:ActiveDynamics}. 

\noindent A comparison of the MSDs for the non-equilibrium hertzian disks at different values $\zeta$ is reported in the SM. Our results for the modified hertzian disks demonstrate that the hertzian force of the EPRs alone cannot generate super-diffusion, while the findings for SFPRs  show that stress propagation alone is also not responsible for that. Thus, it is the combination of the two effects, i.e. the simultaneous presence of the extra force and of the internal elasticity, which gives rise to the anomalous dynamics in the EPRs, as discussed in the next section.

\subsection{Effects of the hertzian field and of the elasticity on the dynamics of EPRs: collective motion}
To corroborate the fact that the persistent hertzian force alone cannot generate super-diffusion, we investigate the correlation  between this force and the displacement of the rings. This analysis is  motivated by previous observations of strong spatial correlation among EPRs, which were found to move ballistically in clusters on intermediate timescales, before that diffusion took place~\cite{Gnan}. This suggests that there might be a characteristic size of clusters over which the force is correlated with the displacement. 

\begin{figure}[h!]
\centering
\includegraphics[width=1.\linewidth]{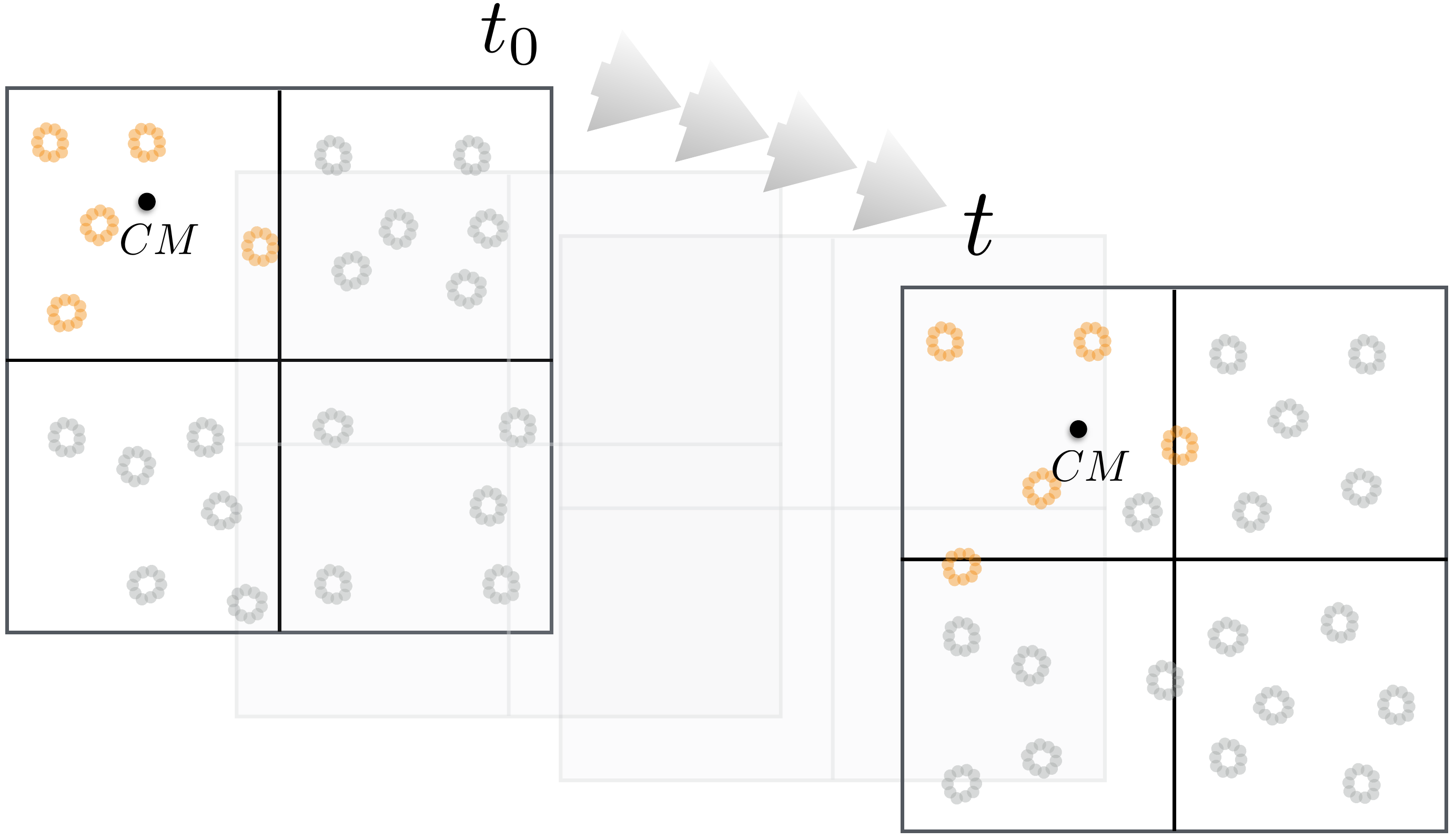}
  \caption{Illustration of the box method described in the text. Initially at $t_{0}$ the simulation box is divided into sub-boxes. Rings in each sub-box are identifed (e.g. orange rings in the upper left sub-box) and the center of mass (black dot) and other quantities of the sub-box evaluated. Although at longer times some of the rings could move outside the initial box and some other could enter in it, all the  quantities evaluated with the method, such as the center of mass, are still calculated considering only the rings belonging to the box at $t_{0}.$}
  \label{fig:boxmethod}
\end{figure}

\begin{figure}[h!]
\centering
\includegraphics[width=1.\linewidth]{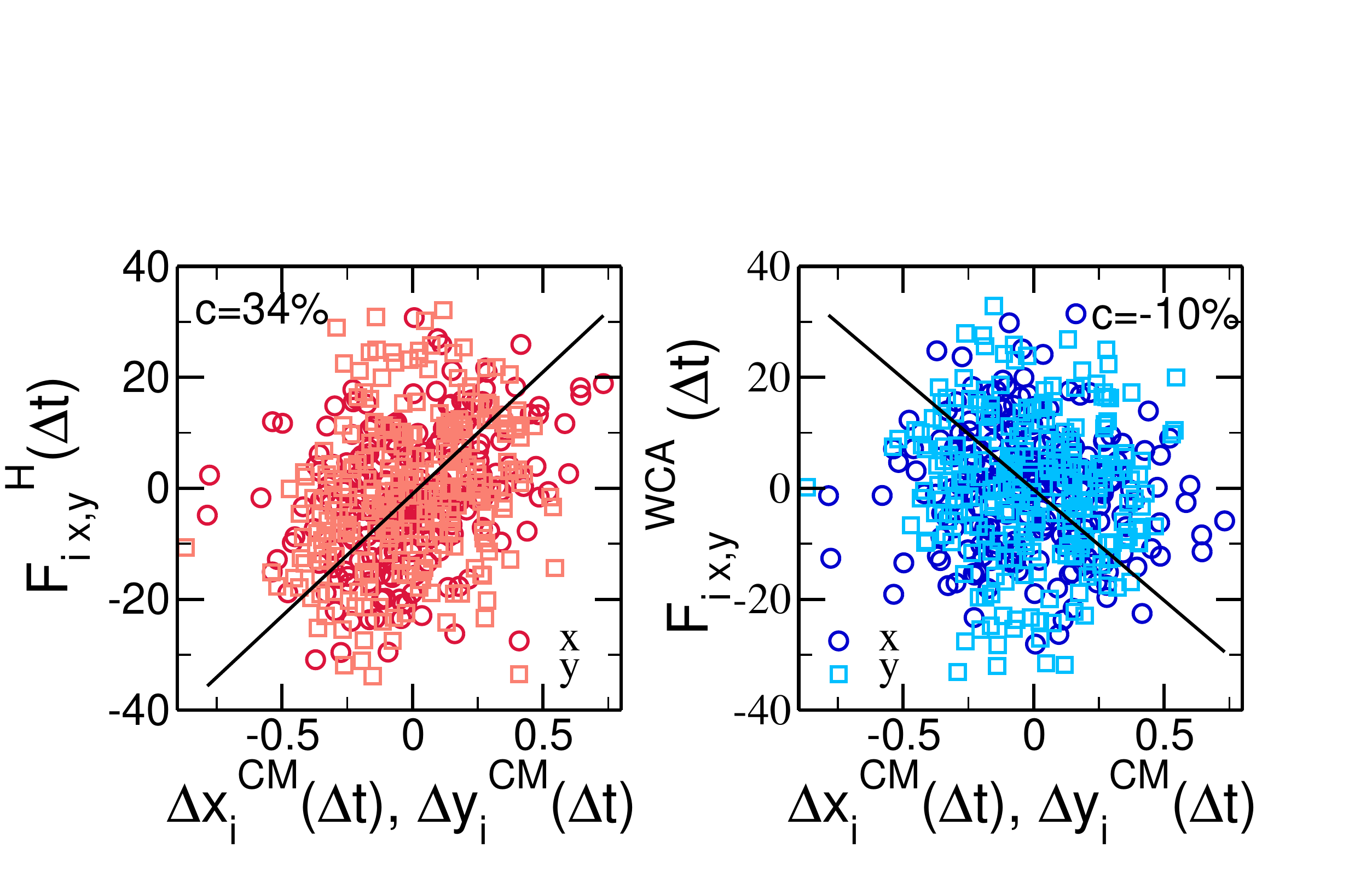}
  \caption{Left panel: x and y components of $F^{H}_{i}$($\Delta t$)  as a function of the displacement of the center of mass $\Delta x^{CM}$($\Delta t$)  and  $\Delta y^{CM}$($\Delta t$) of each sub-box $i$ for 10 windows of length $\Delta t/t_{0}=7.89$.  Right panel: The same as right panel but for $F^{WCA}_{i}$($\Delta t$). The system has been divided into $25$ sub-boxes. Black lines are fits using as slope $m=c\cdot \sigma_{F_{zz}}/\sigma_{\Delta zz}$ where $c$ is the correlation coefficient from linear regression and  $\sigma_{F_{zz}}$, $\sigma_{\Delta zz}$ are standard deviations of $F^{H,WCA}_{i,x}$($\Delta t$) and $\Delta x^{CM}_i$($\Delta t$). Data are for  EPRs with $U=1000$ at $\zeta=1.26$. }
  \label{fig:separateF}
\end{figure}

To this aim, we divide the simulation box into $n_b$ sub-boxes and identify all the particles that, in the initial time, belong to that box. An illustration of the procedure is shown in Fig.~\ref{fig:boxmethod}. Within each sub-box $i$ we calculate the instantaneous total force $\vec{F}^{tot}_{i}(t)=\sum_{j=1}^{N_{i}}\vec{F}^{tot}_j(t)$ and the center of mass  $\vec{R}^{CM}_{i}(t)=\sum_{j=1}^{N_{i}}\vec{r}^{CM}_{j}(t)$ where the index $j$ runs over the $N_i$ rings belonging to the sub-box. Then, we average the force over $M_t$ configurations corresponding to a time window $\Delta  t$, i.e. $\vec{F}^{tot}_{i}(\Delta t)=\frac{1}{M_{t}}\sum_{t=1}^{M_t} \vec{F}^{tot}_{i}(t)$ and calculate the displacement of the center of mass within the the same time window $\Delta \vec{R}^{CM}_{i}(\Delta t)=\vec{R}^{CM}_{i}(t+\Delta t)-\vec{R}^{CM}_{i}(t)$ where we indicate the $(x,y)$ components of the vector as  $\Delta\vec{R}^{CM}_i=(\Delta x^{CM}_i,\Delta y^{CM}_i)$. We then repeat the analysis for several box sizes, aiming to identify the emergence of a characteristic box size.
Fig. \ref{fig:separateF} (left panel) shows the components of $\vec{F}^{H}_{i}(\Delta t)$ of EPRs as a function of the corresponding components of $\Delta \vec{R}^{CM}_{i}(\Delta t)$ for several time windows of  length $\Delta t/t_0=7.89$, roughly coinciding with the characteristic time of super-diffusion~\cite{Gnan}. To obtain the data shown in Fig. \ref{fig:separateF} the  simulation box was divided into $5\times5$ boxes, each of which contains a number of rings compatible with the size of clusters of super-diffusive EPRs observed in Ref.~\protect\cite{Gnan} for the same state point. The figure shows that there is a tiny correlation between the force and the displacement. The same small correlation is found  (with no specific trend) for different number of sub-boxes, down to the limit case in which the force-displacement correlation is investigated at the single-particle level. $\vec{F}^{H}$ is not the only force that acts on the center of mass of a single ring. In fact there is also the contribution of the WCA force $F^{WCA}$ arising from the interaction with neighbor rings. Fig. \ref{fig:separateF} (right panel) shows that also this force, averaged within boxes, is not correlated with the displacement (as it is expected to be). A similar analysis for SFPR is reported in the SM, showing that, for that model, a correlation between the force acting on the center of mass of the box and its displacement is never found.

However, we find that if the contribution $\vec{F}^{tot}=F^{WCA}+F^{H}$ is taken into account, then a strong correlation between the force and the displacement is found.

\begin{figure}[h!]
\centering
\includegraphics[width=1.\linewidth]{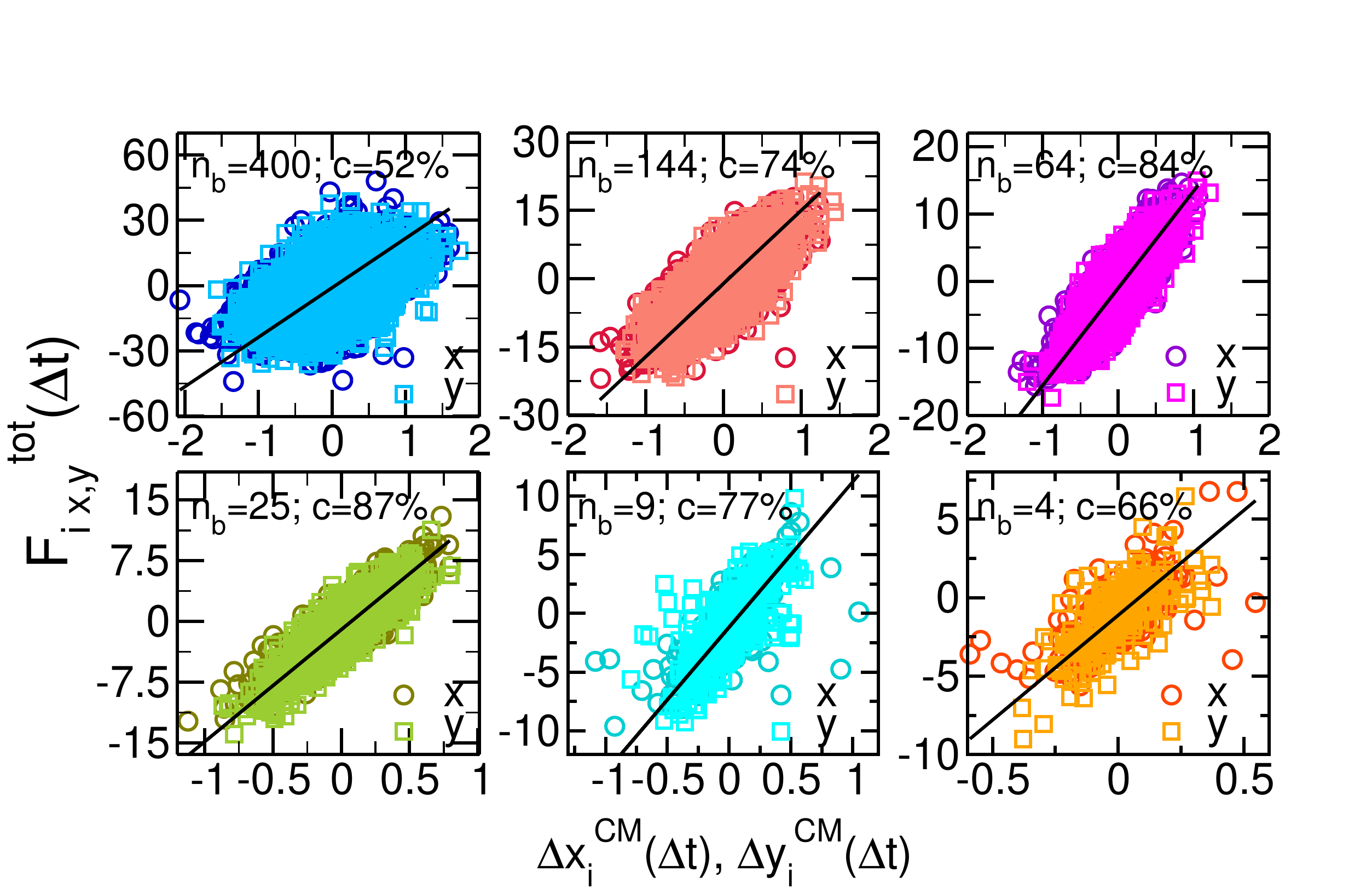}
  \caption{Components $(x,y)$ of $\vec{F}^{tot}_{i}(\Delta t)$  as a function of the  respective components of the displacement of the center of mass $\Delta \vec{R}^{CM}_{i}(\Delta t)$ of each sub-box $i$. Data are taken from trajectories of  EPRs with $U=1000$ at $\zeta=1.26$ over 49 time intervals, each of length $\Delta t/t_{0}=7.89$, that is the characteristic time of super-diffusion~\protect\cite{Gnan}. The analysis is repeated for different sub-box sizes: the simulation box (of total length $\sim 123\sigma_m$ x$123\sigma_m$) is divided into a decreasing number of boxes $n_b$ from top left to bottom right, namely $n_b= 400, 144, 64, 25, 9$ and $4$. From a linear regression of the data, the correlation coefficient $c$ between the force and the displacement components is estimated. Black lines are fits using as slope $m=c\cdot \sigma_{F_{zz}}/\sigma_{\Delta zz}$ where  $\sigma_{F_{zz}}$, $\sigma_{\Delta zz}$ are standard deviations of $F^{tot}_{i,x}$($\Delta t$) and $\Delta x^{CM}_i$($\Delta t$).}
  \label{fig:boxsize}
\end{figure} 

Figure \ref{fig:boxsize} shows the components of $\vec{F}^{tot}_{i}(\Delta t)$ as a function of the corresponding components of $\Delta \vec{R}^{CM}_{i}(\Delta t)$ for several time windows of  length $\Delta t/t_0=7.89$ and for different sizes of the sub-boxes. We notice that there is a characteristic size of the sub-boxes, corresponding to a number of boxes $n_b=25$, for which the correlation  between the force and the displacement is  maximum. Again, such characteristic size contains a number of rings which is consistent with the size of clusters that move collectively  at the considered $\zeta$ and $U$ described in Ref. ~\cite{Gnan}.  
Such results highlight the importance of the interplay between the two forces in generating the collective dynamics. In the SM we also show that, in the case of the modified hertzian disks, the correlation between  $\vec{F}^{tot}$ or $\vec{F}^{H}$ with the displacement does not exist at any sub-box subdivision.

In summary, we observe a strong interplay between  excluded volume interactions and the hertzian force, which gives rise to a motion that is spatially correlated.  Recent works have shown that simple models of active particles, in which the alignment of the velocity  is not introduced {\it ad hoc} in the interaction rules, can  also display a coherent motion due to the action of elastic interactions~\cite{Ferrante2013}. Our system at high packing fraction also displays an alignment of the displacement and the underlying elastic interactions are connected to such coherent motion.  However, differently from simple active models, the force generated by the hertzian field in our model does not follow its own independent evolution, but it is self-generated through the mechanical stress that arises within each EPR, therefore it is difficult to predict what would be its effect on the dynamics of the system.  Further work on this issue will be needed in the future, which should aim to the development of simpler models with features similar to the present EPRs as well as clearer connections to existing active models.

\section{Discussion and Conclusions}
In the present work we compared three different models of polymer rings in order to understand what are the general features in the dynamical behavior of a simple model of particles with an internal elasticity. This work largely extends our previous study~\cite{Gnan} where the EPR system was investigated and a reentrant dynamics was detected, being characterized by a peculiar super-diffusive behavior over intermediate timescales. In such a work, the non-equilibrium features of the EPR model, that arise due to the presence of the inner hertzian field, were overlooked{$^{32}$}. 
Here, we fully unveil these non-equilibrium features, by explicitly discussing the role played by the hertzian field, which can be considered equivalent to an active force, self-generating due to the deformation of polymer rings. The contribution of such a force is negligible when particle deformation occurs due to thermal fluctuations and the system behaves as if it were in equilibrium. However, at higher packing fractions, excluded volume effects give rise to {asymmetric} deformations which generate a non-zero  persistent force  acting on each ring. By investigating the dynamical behavior of three models of polymer rings, we establish a clear link between deformation and fragility, as previously determined in Ref.~\cite{Gnan} for the EPR system only. To this aim, we investigated the dependence of the fragility on the softness of rings, quantified by the average asphericity, finding that, indeed, the link between these two quantities is a generic feature of elastic, deformable particles. These results show that there exists a direct connection between the microscopic elastic properties of the particles and their dynamical behavior, which is expected to hold also in 3D and for more refined models. We aim to further elucidate this aspect with more realistic models in the near future. 

Regarding the super-diffusive dynamics, we attempted to provide a direct evidence of the hypothesized connection between the release of stress within the system and the occurrence of the anomalous dynamics. To this aim, we compared the MSDs of the three different models of polymer rings, finding that super-diffusion at intermediate time-scales only occurs for EPRs, but not for eq-EPRs and SFPRs.  By analyzing the ability of the three kinds of rings, not only to deform, but also to fluctuate in shape over time, we also provided evidence that the semi-flexible model has the greatest ability to both deform and fluctuate, but again without showing super-diffusion. However, the stress to be released in the SFPR model is {significantly} smaller than the corresponding one for the EPRs, where the active force pumps energy into the system at all times for high $\zeta$. We therefore speculate that another mechanism is at work in the EPR system which is able to { induce stress propagation among neighbor rings, giving rise to a (coherent) super-diffusive dynamics}. This mechanism does not owe only to the presence of the out-of-equilibrium force, but is also related to the presence of internal elasticity. In fact, we find that the out-of-equilibrium force is not directly correlated to rings displacement as one could initially guess. Instead, we find that the activity is mediated by excluded volume interactions originating a motion that is spatially correlated over specific length scales that depend on $\zeta$ and on the model parameters (i.e. ring softness). The displacement of the regions of coherent rings turns out to be highly correlated with the total force, i.e. the sum of the WCA force and the inner hertzian force, but not with the two contributions separately. This strongly suggests that ring elasticity plays a crucial role for the emergence of such coherent motion, as also observed in simple active models~\cite{Ferrante2013}. 
The fact that the latter  is a key ingredient to obtain super-diffusion was directly proven by building an alternative system, the so-called modified hertzian disks, where a self-generated persistent force based on the overlap among particles was introduced. Despite this force being highly correlated in time as in the case of EPRs, no super-diffusion was obtained, because of the lack of internal elasticity of the particles. Our work thus suggests that anomalous dynamics must be linked to out-of-equilibrium features, which act in combination with other microscopic ingredients, such as internal elasticity of soft particles. Hence, within the present study, it appears that for purely equilibrium systems a faster-than-diffusive (or exponential) dynamics cannot be observed. These findings are in agreement with recent experimental results~\cite{Dallari2020}, which showed that a faster-than-exponential relaxation in colloidal glasses was related to the existence of a pre-stress condition in the samples. This in turn originates the out-of-equilibrium dynamics that is necessary for the occurrence of the anomalous dynamics.  Notwithstanding this, several questions remain open related to {stress relaxation and stress propagation in these systems at high densities. In particular, the present results do not allow us to relate the onset of the anomalous dynamics to the intensity of the released stress, because the SFPR model never reaches values of total stress as high as those of the EPRs. In the future, it would be interesting to study some similar models, where this aspect could be tested in more detail. In addition, it would be important to design simpler non-equilibrium models, possibly amenable of theoretical treatment, that could help us to shed light on the exact mechanism occurring in EPR  leading to stress propagation and to the emergence of anomalous dynamics.}

\section*{Supplementary Material}
\noindent See supplementary material for a discussion on the choice of parameters for the EPRs and the eq-EPRs, for additional results on the modified hertzian disks and on the SFPRs.

\begin{acknowledgments}
The authors acknowledge financial support from the European Research Council (ERC Consolidator Grant 681597, MIMIC) and from MIUR (FARE project R16XLE2X3L, SOFTART).
\end{acknowledgments}
\section*{Data availability}
The data that support the findings of this study are available from the corresponding author upon reasonable request.

\bibliography{draft}
\clearpage
\newpage
\onecolumngrid
\setcounter{section}{0}
\setcounter{figure}{0}  
\titleformat*{\section}{\LARGE\bfseries}
\section*{Supplementary Material of Dynamical properties of different models of elastic polymer rings: confirming the link between deformation and fragility}
\titleformat*{\section}{\normalfont\bfseries}
\section{Choice of parameters for EPR and eq-EPR models.}
\begin{figure}[!h]
\centering
\includegraphics[width=0.8\linewidth]{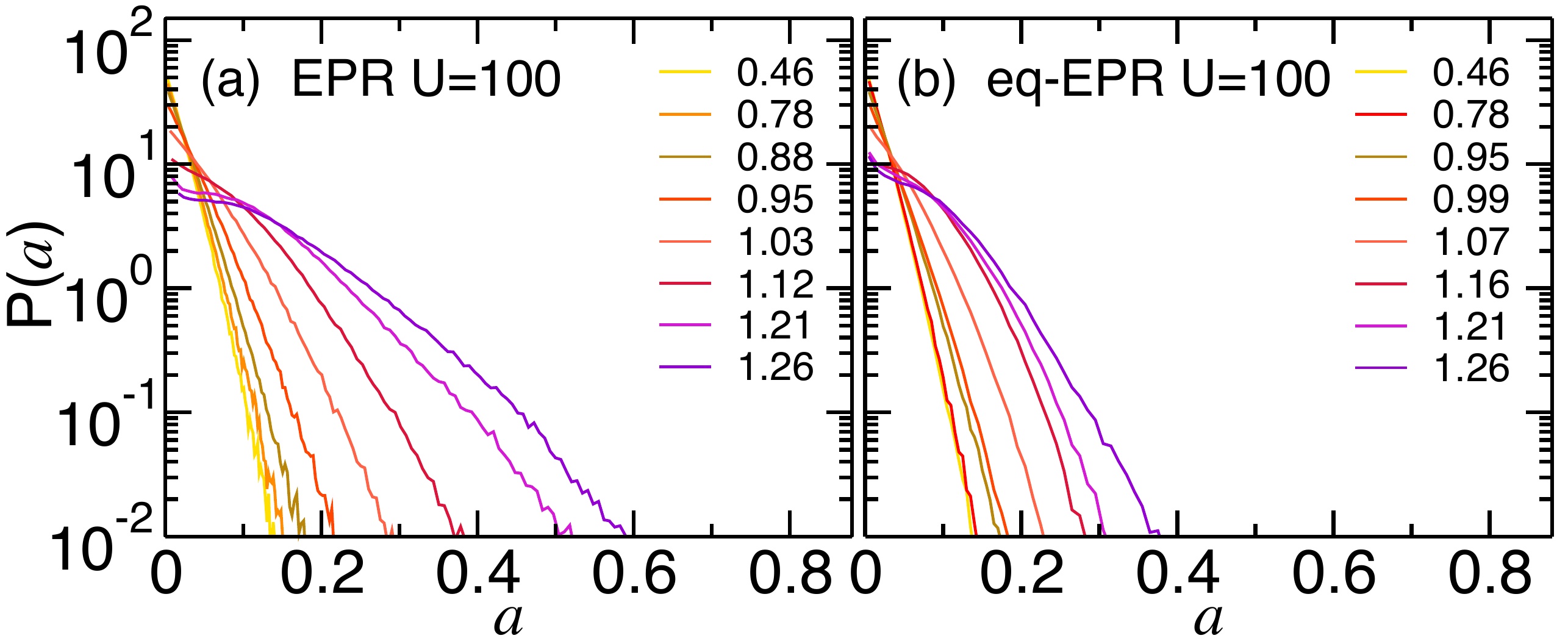}
\caption{\label{fig:AsphCompareU100} 
Asphericity distribution for (a) EPR with $U=100$ and (b) eq-EPR with $U=100$.}
\end{figure}

As discussed in the main text the overall effect  of the balancing force $\vec{f}_{CM}$ in the eq-EPR is to reduce rings deformation at high packing fractions. This can be observed e.g., by comparing the asphericity distribution of EPRs and eq-EPRs at the same $U=100$ as shown in Fig.~\ref{fig:AsphCompareU100}. Notice that at the same $U$ and low $\zeta$, there is no difference in the asphericity between two models, while deviations occur when rings start to be strongly in contact with each other

\begin{figure}[!h]
\centering
\includegraphics[width=0.8\linewidth]{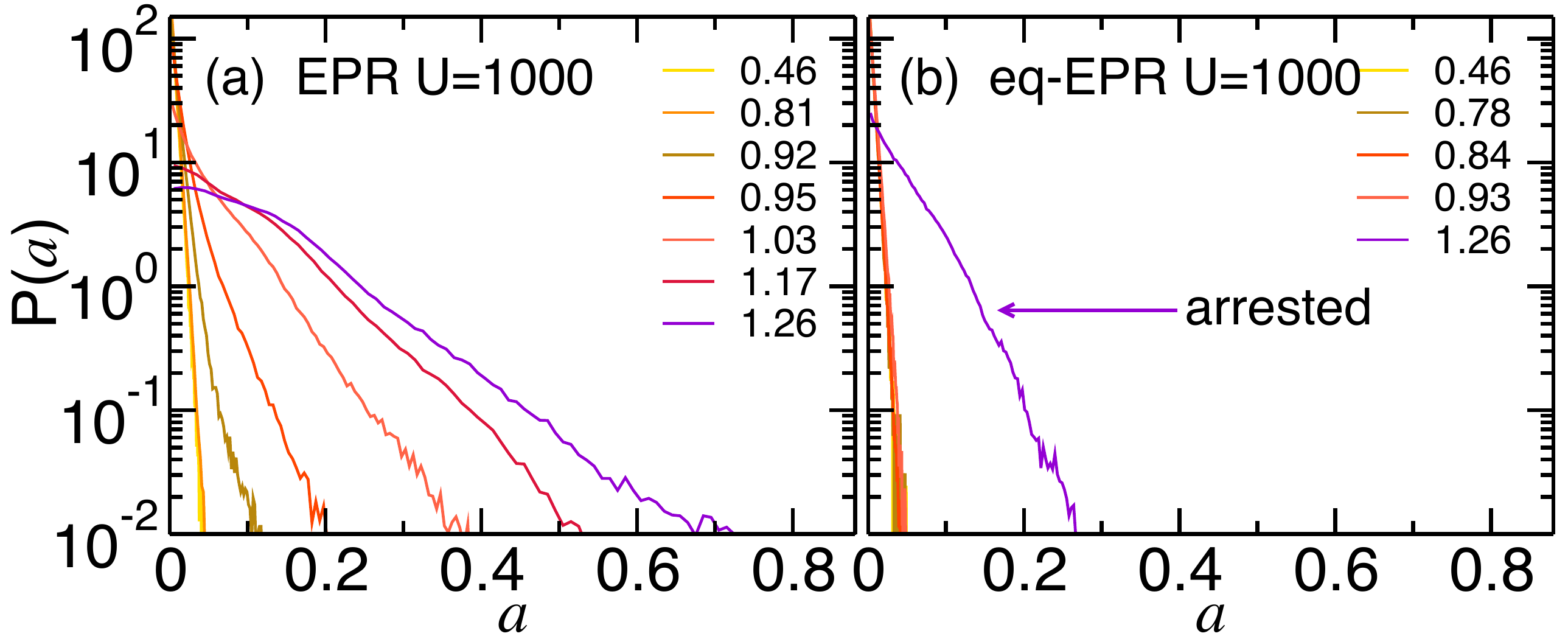}
\caption{\label{fig:AsphCompareU1000} 
Asphericity distribution for (a) EPR with $U=1000$ and (b) eq-EPR with $U=1000$.}
\end{figure}

It is interesting to notice that such difference becomes quite large by increasing $U$ as shown for instance in Fig.~\ref{fig:AsphCompareU1000} where the same comparison in displayed for $U=1000$.

\begin{figure}[!h]
\centering
\includegraphics[width=0.8\linewidth]{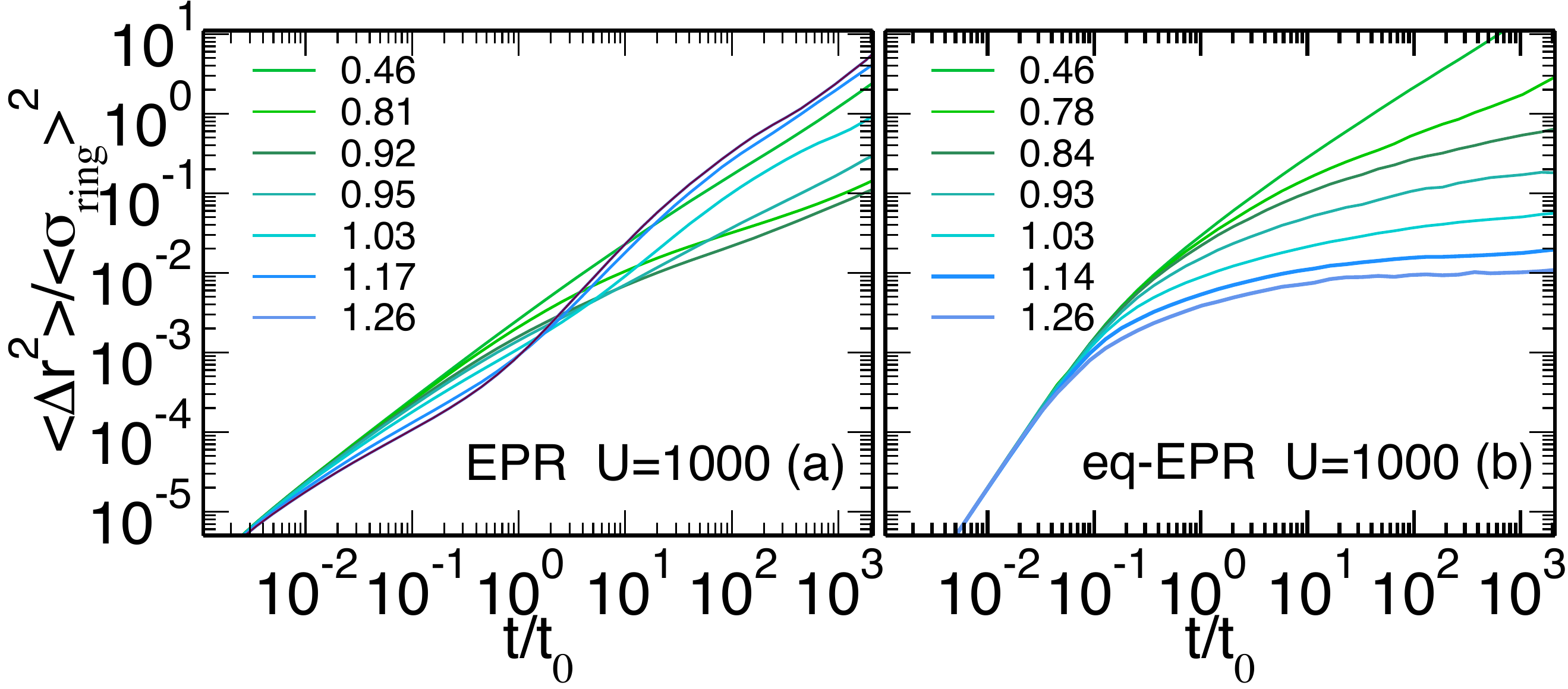}
\caption{\label{fig:MSDCompareU1000} 
MSD for (a) EPR with $U=1000$ and $D_0=0.008$ and (b) eq-EPR with U=1000 and $D_0=0.08$.}
\end{figure}

Fig.~\ref{fig:AsphCompareU1000}  clearly shows that the asphericity distribution of the eq-EPR does not change on increasing $\zeta$ up to the jamming point. Beyond jamming, initial configurations of the eq-EPR  are taken from final runs of the EPR model at the same $\zeta$. Differently from the EPR the eq-EPR is totally arrested and the  asphericity distribution depends on the initial configuration, i.e. it is not a distribution of an equilibrated system (the initial configuration of the run was taken form the final run of EPRs at the same $\zeta$). To observe the striking difference between the dynamical behaviour of the EPR  and eq-EPR at $U=1000$ we show in Fig.~\ref{fig:MSDCompareU1000} the mean-squared displacement (MSD) of the two models: the eq-EPR MSD does not show any reentrance, and beyond close packing, the system is totally arrested as if rings were almost like hard-disks.

\noindent 
Therefore, for the eq-EPR, the reentrance occurs only in a small U-range and for higher values (beyond $U=200$) the systems hardens, the reentrance disappears and we cannot probe high density states in equilibrium. This is why it was not possible to exploit the same U range to compare the two models. It is important to stress that, independently on the U value chosen for the eq-EPR, a superdiffusive regime has never been observed in the eq-EPR model.

\section{Modified Hertzian disks to understand the role of non-equilibrium effects in the superdiffusive regime of EPR.}
\begin{figure}[h!]
\centering
\includegraphics[width=0.5\linewidth]{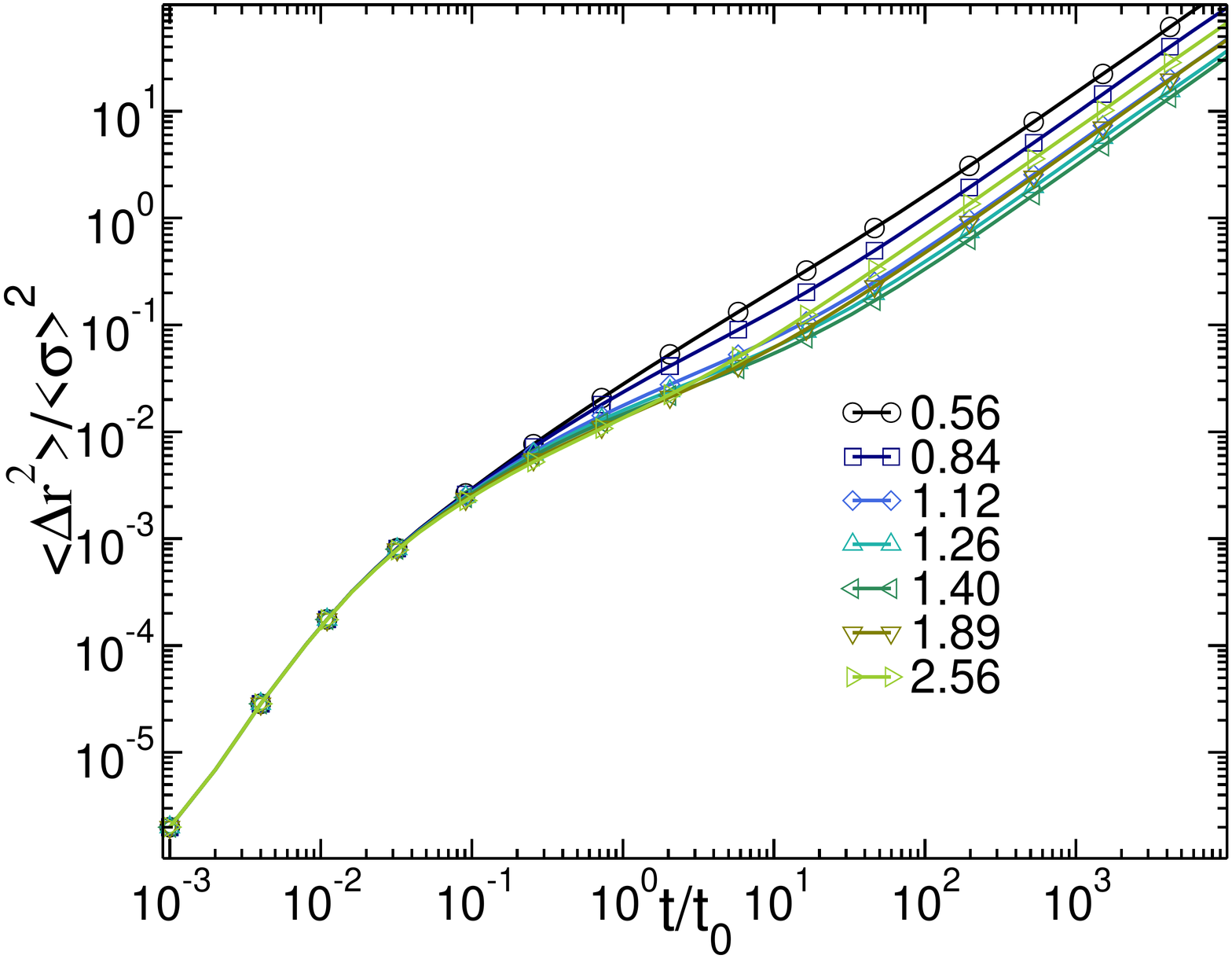}
  \caption{MSD of modified hertzian disks  with $U=150$  and $K=150$ at different $\zeta$ values.}
  \label{fig:ActiveDisksPhiVal}
  \end{figure}

To get more insight into the role played by the Hertzian force in the superdiffusive behaviour of the system, we have proposed a model of soft disks (no internal degrees of freedom), called modified hertzian disks, which have roughly the same features of the the EPR model. Hertzian disks interact with a standard hertzian potential with amplitude $U_{H}=150$ complemented by a  force which depends on the overlap between two disks and a prefactor $K$ (see main text for the description). Since the conservation of the overlap force is not guaranteed due to the polydispersity of the disks,  it can be considered as an off-equilibrium, active force as in the case of the EPRs. The overlap force,  has the effect of pushing disks closer thus increasing the available volume at disposal  for particles and speeding up the dynamics (in analogy to what observed also for the EPR if compared with the eq-EPR). In addition, the system shows a reentrant behaviour. Despite the analogy with the EPRs, the proposed system never displays superdiffusion at any timescale as shown in Fig.~\ref{fig:ActiveDisksPhiVal} where the MSDs of modified hertzian disks are shown for different values of $\zeta$.

\begin{figure}[h!]
\centering
\includegraphics[width=0.5\linewidth]{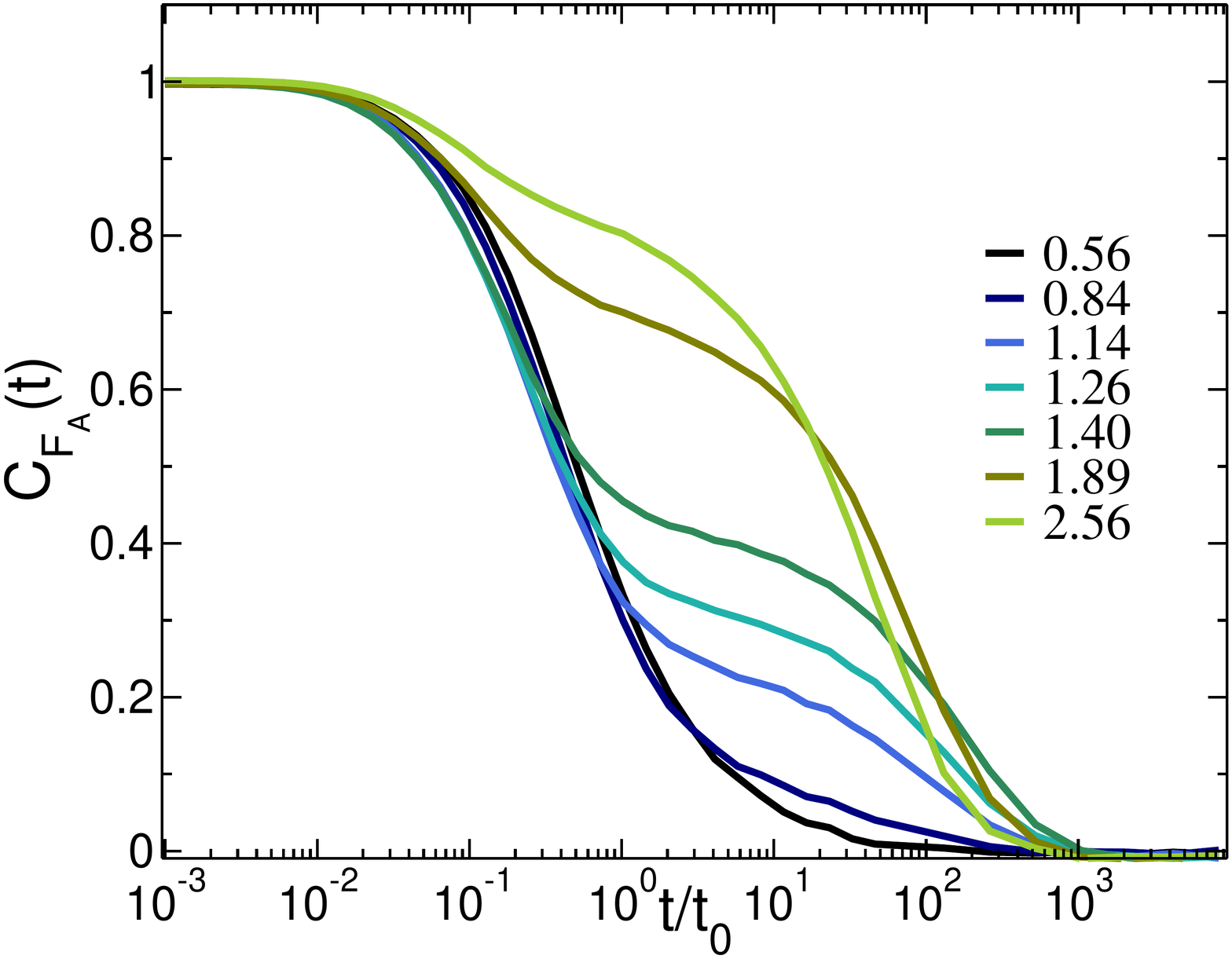}
  \caption{Autocorrelation of the $x$ component of the non-equilibrium force force  $C_{F^{A}}(t)$ of disks with  $U_{H}=150$ and K=$150$ at different packing fractions (see the legend).}
  \label{fig:CorrForceDisks}
  \end{figure}

 We also show in Fig.~\ref{fig:CorrForceDisks}  the correlation function ${C_{F^{A}}}(t)$  calculated for the modified hertzian disks with $U_{H}=150$ and K=$150$ at several packing fractions. Note that, especially at $\zeta=2.56$ (reentrant point) the correlation resembles that of the EPRs; however  this is not a sufficient condition to trigger any kind of anomalous dynamics in the system.
  
\begin{figure}[h!]
\centering
\includegraphics[width=0.5\linewidth]{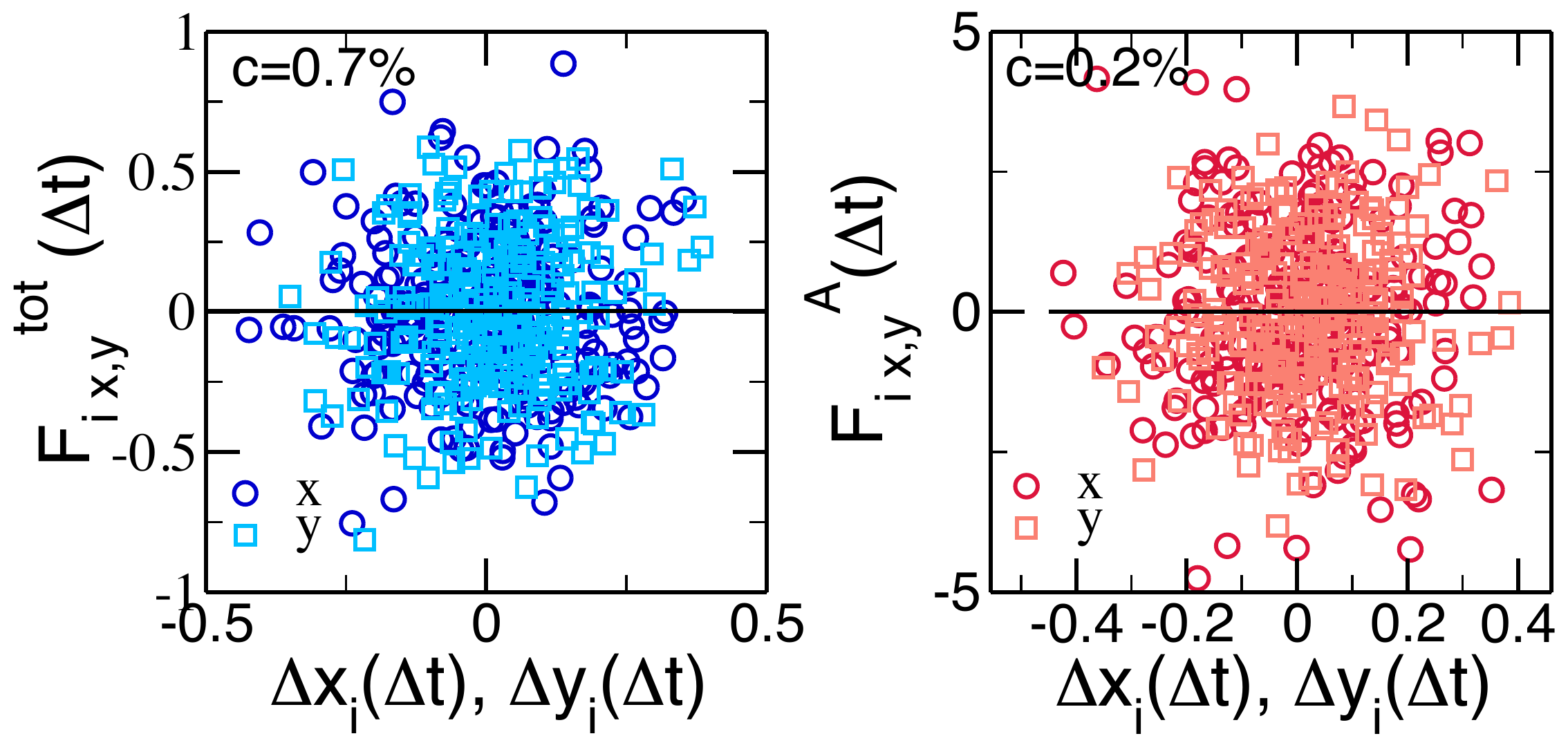}
  \caption{Correlation between forces and displacements in the modified Hertzian model. Left panel: x and y components of $F^{tot}_{i}$($\Delta t$)  as a function of the displacement  $\Delta x_i$($\Delta t$)  and  $\Delta y_i$($\Delta t$) of each sub-box $i$ for 10 windows of length $\Delta t/t_{0}=7.89$.  Right: The same as left panel but for $F^{A}_{i}$($\Delta t$).The system has been divided into $n_b=25$ sub-boxes. Black lines are fits using as slope $m=c\cdot \sigma_{F^{zz}}/\sigma_{\Delta zz}$ where $c$ is the correlation coefficient obtained from linear regression and  $\sigma_{F^{zz}}$, $\sigma_{\Delta_{zz}}$ are standard deviations of $F^{A,tot}_{i,x}$($\Delta t$) and $\Delta x_i$($\Delta t$).
  Data refer to disks with $U_{H}=150$,$K=150$ and $\zeta=2.56$, i.e. in the reentrant region.}
  \label{fig:Box5_totDISKS}
  \end{figure}

Finally, we applied the box method described in the main text to understand whether a correlation exists between the forces acting on the disks and the disks displacement.
Fig.~\ref{fig:Box5_totDISKS}  shows the box analysis for the modified hertzian disks with $U_{H}=150$ and K=$150$ at $\zeta=2.56$ (reentrant point). Both the total force $F^{tot}=F^A + F^H$ (left panel) and the active force $F^{A}$ only (right panel) are considered. The simulation box has been divided into into $n_b=25$ sub-boxes and forces have been time averaged within a time window $\Delta t/t_{0}=7.89$ to compare results with those of the EPR system. For both forces, no correlation is found.

\begin{figure}[h!]
\centering
\includegraphics[width=0.5\linewidth]{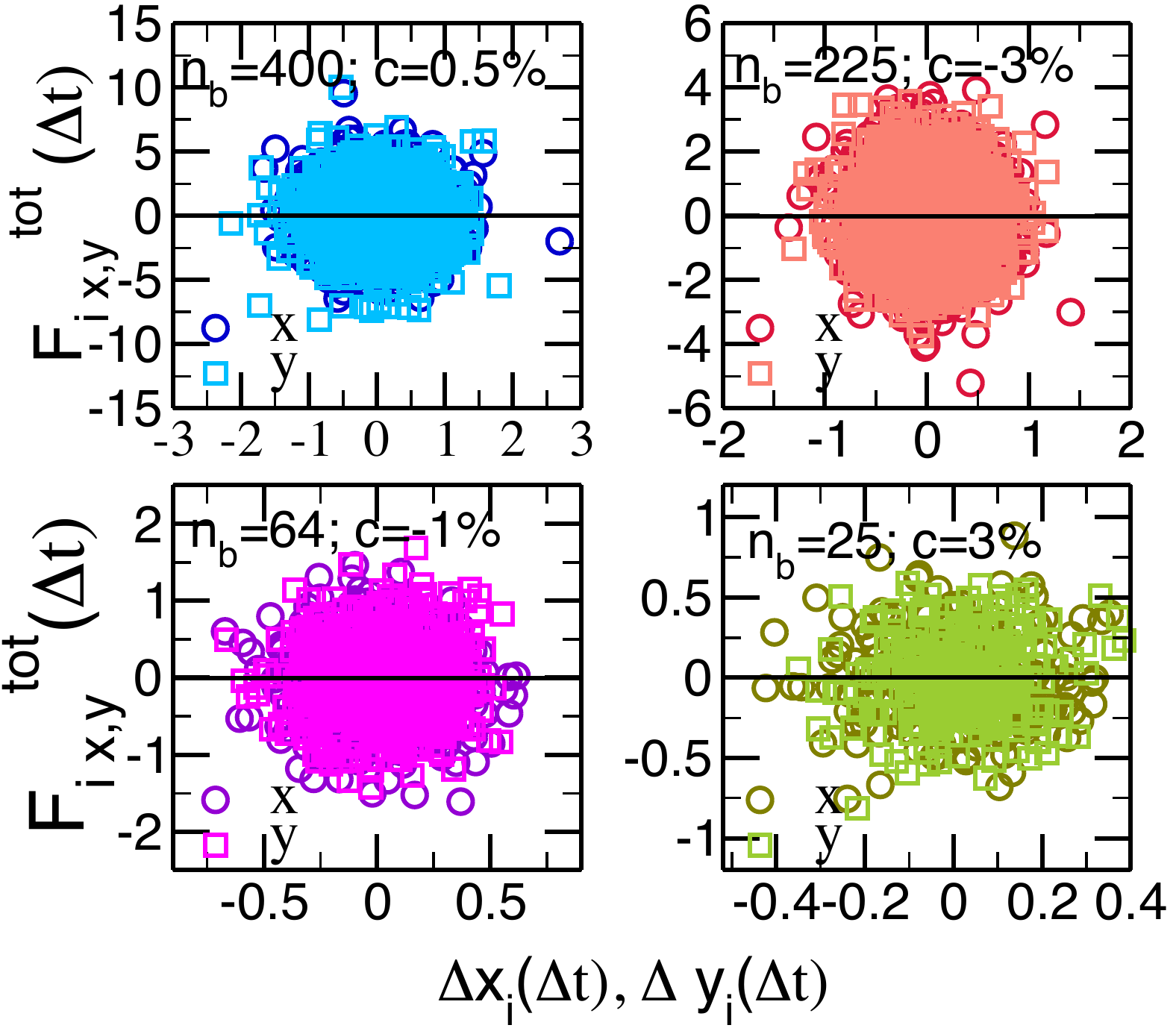}
\caption{x and y components of $F^{tot}_{i}$  as a function of the  components of the displacement of the center of mass of each sub-box $i$ for a single window of length $t/t_0=7.89$. Data refer to the modifed hertzian disks   with $U_{H}=150$,$K=150$ and $\zeta=2.56$, i.e. in the reentrant region. Black lines are fits using as slope $m=c\cdot \sigma_{F^{zz}}/\sigma_{\Delta zz}$ where $c$ is the correlation coefficient obtained from linear regression and  $\sigma_{F^{zz}}$, $\sigma_{\Delta zz}$ are standard deviations of $F^{tot}_{i,x}$($\Delta t$) and $\Delta x_i$($\Delta t$).}
\label{fig:Box20_totDISKS}
\end{figure}

We also show the dependence on the box subdivision of the correlation of the total force in Fig.~\ref{fig:Box20_totDISKS}. Again no correlation is present.

\section{Box analysis for semi-flexible polymer rings}
\begin{figure}[h!]
\centering
\includegraphics[width=0.5\linewidth]{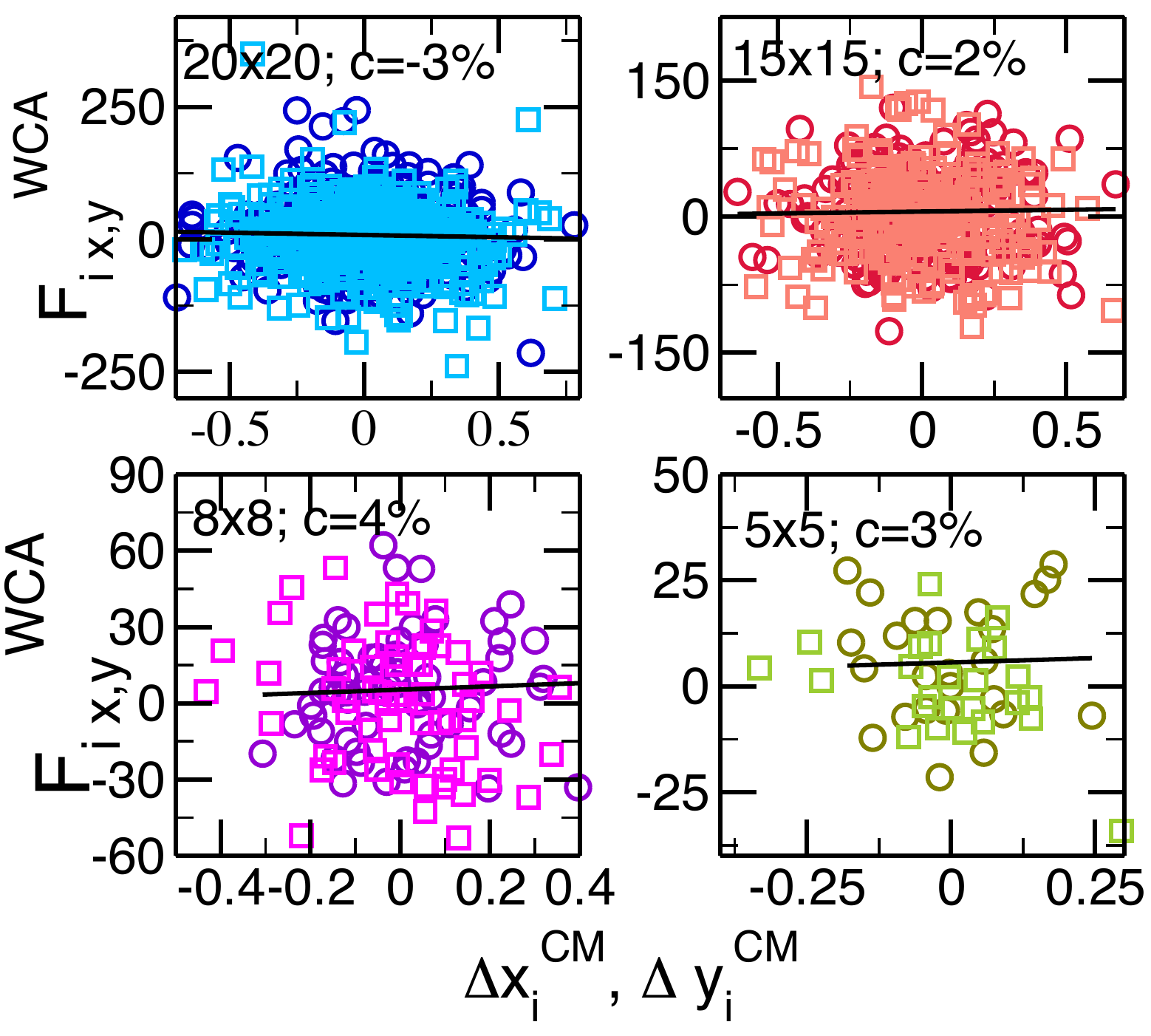}
  \caption{x and y components of $F^{WCA}_{i}$  as a function of the  components of the displacement of the center of mass of each sub-box $i$ for a single window of length $t/t_0=7.89$. Data refer to the SFPR  with $k_{\theta}=5$ at $\zeta=1.26$. Black lines are fits using as slope $m=c\cdot \sigma_{F^{zz}}/\sigma_{\Delta zz}$ where $c$ is the correlation coefficient obtained from linear regression and  $\sigma_{F^{zz}}$, $\sigma_{\Delta zz}$ are standard deviations of $F^{WCA}_{i,x}$($\Delta t$) and $\Delta x_i$($\Delta t$).}
  \label{fig:boxsize_bsa_1.26}
\end{figure}

\begin{figure}[h!]
\centering
\includegraphics[width=0.5\linewidth]{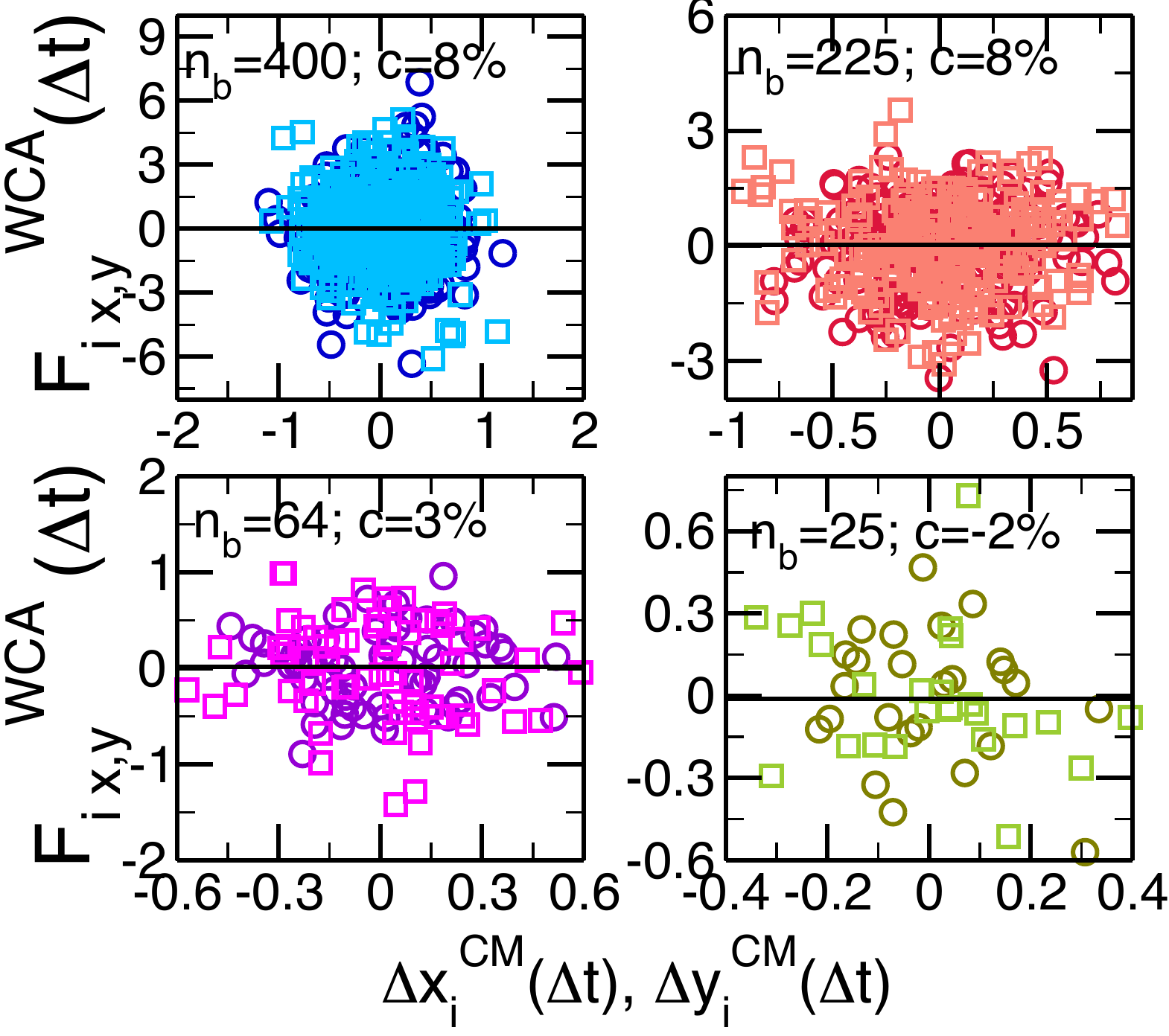}
  \caption{x and y components of $F^{WCA}_{i}$  as a function of the  components of the displacement of the center of mass of each sub-box $i$ for a single window of length $t/t_0=7.89$. Data refer to the SFPR  with $k_{\theta}=5$ at $\zeta=1.01$ (reentrant point). Black lines are fits using as slope $m=c\cdot \sigma_{F^{zz}}/\sigma_{\Delta zz}$ where $c$ is the correlation coefficient obtained from linear regression and  $\sigma_{F^{zz}}$, $\sigma_{\Delta zz}$ are standard deviations of $F^{WCA}_{i,x}$($\Delta t$) and $\Delta x_i$($\Delta t$).}
  \label{fig:boxsize_bsa_1.0135}
\end{figure}

Here, we show the box analysis proposed in the main text for EPRs, this time applied on the SFPR model with $k_{\theta}=5$. In this case the only force acting on the center of mass is that of the WCA coming from collisions with other rings in the simulation box.
We concentrate on two different packing fraction: $\zeta=1.26$ to compare with the EPR analysis and $\zeta=1.01$ i.e. within the region of the reentrant dynamics.
For these sets we have a single time window of $\Delta t/t_{0}=7.89$ to observe if there is a correlation between force and displacement. Results for high and low $\zeta$ at different box sizes are shown, respectively, in Fig.~\ref{fig:boxsize_bsa_1.26} and Fig.~\ref{fig:boxsize_bsa_1.0135} (i.e. in the reentrant region). In both cases we find that no correlation between motion and force, independently from the spatial average performed. 

\end{document}